\documentclass[aps,prl,10pt,twocolumn,twoside,preprintnumbers,superscriptaddress, 
amsmath,
amssymb,
flushbottom]{revtex4-2}

\usepackage{physics}
\usepackage{hyperref}
\usepackage{enumitem}
\usepackage{xspace}
\usepackage{slashed}
\usepackage{braket}
\usepackage{leftindex}
\usepackage{graphicx}  
\usepackage{bm}  
\usepackage{booktabs}
\usepackage[normalem]{ulem}
\usepackage{graphics}
\usepackage{color}
\usepackage{colortbl}
\usepackage[svgnames,dvipsnames,table,x11names]{xcolor}
\usepackage{comment}
\usepackage{enumitem}
\usepackage{mathrsfs}

\hypersetup{
    pdfencoding=unicode,
	colorlinks=true,
	urlcolor=RoyalBlue,
	linkcolor=Maroon,
	citecolor=ForestGreen
}

\newcommand{\agl}[2]{\langle #1 #2 \rangle}
\newcommand{\sqr}[2]{\lbrack #1 #2 \rbrack}
\newcommand{\Mpl}[0]{M_{\text{P}}}
\newcommand{\hf}[0]{\tfrac{1}{2}}

\makeatletter\g@addto@macro\bfseries{\boldmath}\makeatother

\begin{document}

\allowdisplaybreaks

\title{Amplitudes and partial wave unitarity bounds}

\author{Luigi C.~Bresciani}
\email{luigicarlo.bresciani@phd.unipd.it}
\affiliation{Dipartimento di Fisica e Astronomia ``G.~Galilei'', Università degli Studi di Padova, Via F.~Marzolo 8,
35131 Padova, Italy}
\affiliation{Istituto Nazionale di Fisica Nucleare, Sezione di Padova, Via F.~Marzolo 8, 35131 Padova, Italy}

\author{Gabriele Levati}
\email{gabriele.levati@unibe.ch}
\affiliation{Albert Einstein Center for Fundamental Physics, Institute for Theoretical Physics, University of Bern, Sidlerstrasse 5, 3012 Bern, Switzerland}

\author{Paride Paradisi}
\email{paride.paradisi@unipd.it}
\affiliation{Dipartimento di Fisica e Astronomia ``G.~Galilei'', Università degli Studi di Padova, Via F.~Marzolo 8,
35131 Padova, Italy}
\affiliation{Istituto Nazionale di Fisica Nucleare, Sezione di Padova, Via F.~Marzolo 8, 35131 Padova, Italy}

\begin{abstract}
We develop a formalism, based on spinor-helicity techniques, to generalize the formulation of partial wave unitarity bounds. We discuss unitarity bounds for $N \to M$ (with $N,M \geq 2$) scattering processes---relevant for high-energy future colliders---and spin-2 or higher-spin theories---relevant for effective field theories of gravity---that are not approachable by standard methods. Moreover, we emphasize the power and complementarity of positivity and partial wave unitarity bounds to constrain the parameter space of effective field theories.
\end{abstract}

\maketitle

\paragraph{Introduction.}
New interactions of nature can be probed through a wide variety of experimental tests, encompassing direct searches of new particles at high-energy colliders, indirect signals in low-energy precision measurements, as well as cosmological observations. 
On the theoretical side, perturbative unitarity---relying on the properties of the scattering $S$-matrix---provides an independent and complementary way to infer upper limits on the strength of such interactions, irrespectively of any experimental information.
Generally, unitarity violation signals the breakdown of the low-energy description of a theory which can be restored by introducing new degrees of freedom or a new strong dynamics at the scale where unitarity is broken.
In the past, unitarity bounds for the $WW$ scattering process led to the famous no-lose Higgs theorem, which implied an upper bound on the Higgs boson mass below the TeV scale~\cite{Lee:1977yc,Lee:1977eg}.
Such a theoretical prediction was of paramount importance in motivating the construction of the LHC at CERN. 

More recently, unitarity bounds have been extensively discussed in the literature to assess the range of validity of Effective Field Theories (EFTs)~\cite{Gounaris:1994cm,Corbett:2014ora,Corbett:2017qgl,DiLuzio:2017chi}.
Indeed, non-renormalizable interactions generate scattering amplitudes growing with energy, thus leading to unitarity violation at some large energy scale. 
As a result, a correct interpretation of experimental data, such as those inherent to tails of kinematical distributions, has to be necessarily supplemented by unitarity bounds.

The standard way of implementing unitarity bounds relies on $2\to 2$ scatterings of particles with helicities $h_i$ and proceeds in two steps: i) expansion of the helicity amplitudes into partial waves $a_{h_i}^J$ with fixed total angular momentum $J$ by means of the Wigner rotation matrix~\cite{Jacob:1959at} and ii) diagonalization of the partial wave scattering matrix.
In this coupled channel analysis, the most stringent limit is then obtained by imposing the 
partial wave unitarity bound $|a_{h_i}^J|_{\rm max} \lesssim 1$ on the largest eigenvalue.

Although very popular, the above method suffers from two serious shortcomings which significantly reduce its regime of applicability. On the one hand, partial wave decompositions are known only for $2\to 2$ processes. 
However, since in many cases $2 \to N$ (with $N > 2$) amplitudes show a faster growth with energy than $2 \to 2$ amplitudes, unitarity bounds from $2 \to N$ processes
are expected to dominate at high energies, which can be probed at future colliders such as FCC-ee~\cite{FCC:2018byv} or a high-energy Muon Collider~\cite{Accettura:2023ked}.

On the other hand, even in the $2\to 2$ case, a standard partial wave decomposition for spin-2 or higher-spin theories---relevant for EFTs of gravity---is impracticable given the difficulty in evaluating the amplitudes via Feynman rules.
In this respect, on-shell methods---which proved to be very efficient to capture ultraviolet quantum effects of EFTs~\cite{Caron-Huot:2016cwu,EliasMiro:2020tdv,Baratella:2020lzz,Jiang:2020mhe,Bern:2020ikv,Baratella:2020dvw,AccettulliHuber:2021uoa,EliasMiro:2021jgu,Baratella:2022nog,Machado:2022ozb,Bresciani:2023jsu,Bresciani:2024shu,Aebischer:2025zxg}---seem to be the ideal tool also to probe possible unitarity violations of EFTs at high energies.

The primary goal of this Letter is to generalize the formulation of partial wave unitarity bounds in order to overcome the above shortcomings. 
In particular, we elaborate on a vectorial formalism~\cite{Jiang:2020rwz}, based on the spinor-helicity formalism~\cite{Dixon:2013uaa}, which
will allow us to determine the amplitude basis thanks to the amplitude-operator correspondence~\cite{Shadmi:2018xan,Durieux:2019eor,Dong:2021vxo,Li:2022tec}. 
As a result, we find the generalization of the Wigner rotation matrix entering the partial wave expansion of $N \to M$ scattering amplitudes. 
This will enable us to set the most stringent partial wave unitarity bounds on those EFT interactions that generate contact $2\to 3$ scattering amplitudes.

Furthermore, working in a Minkowski background, we discuss partial wave unitarity 
bounds for EFTs of gravity and higher-spin theories, which have been conjectured 
to be connected to weakly coupled conformal field theories through the AdS/CFT correspondence~\cite{Klebanov:2002ja}. The recent direct observation of gravitational waves from merging black holes opens the possibility of probing the theory of gravity in the strong regime at an unprecedented level. It is therefore very interesting to explore which EFT extensions of gravity could be detected. In this respect, our work serves as an important guide for exploring viable models that are consistent with first principles of quantum field theory.

As an interesting byproduct of our study, we analyze for the first time the complementarity and interplay of positivity bounds~\cite{Adams:2006sv}---stemming from the analyticity and causality of scattering amplitudes---and partial wave unitarity bounds. The resulting constraints will be of utmost importance to connect experimental data at colliders to fundamental properties of EFTs. Examples include the possibility to 
use the above bounds as a prior in statistical interpretations and to design phenomenological studies to test positivity and partial wave unitarity bounds at colliders.

\medskip
\paragraph{Generalized partial wave unitarity bounds.}
The partial wave analysis allows one to project 
a generic amplitude $\ket{\mathcal A_{i\to f}}$ 
onto a kinematic basis $\ket{\mathcal B^J_{i\to f}}$
with definite angular momentum $J$ as follows
\begin{equation}
    \ket{\mathcal A_{i\to f}} = \sum_J a^J_{i\to f} \ket{\mathcal B^J_{i\to f}}\,,
    \label{eq:pwd}
\end{equation}
where $a^J_{i\to f}$ are the partial wave coefficients.
The fundamental building blocks of such a decomposition are the Poincaré Clebsch-Gordan coefficients $\mathcal C_{\mathcal I \to *}^{J,h}$ defined as~\cite{Jiang:2020rwz}
\begin{equation}
    \langle{P,J,h|\mathcal I}\rangle = \mathcal C_{\mathcal I \to *}^{J,h}\,
    \delta^{(4)}\!\left(P-\sum_{i\in\mathcal I}p_i\right)\!\,,
\end{equation}
namely the overlap between the multiparticle state $\ket{\mathcal I}$ and the Poincaré irreducible multiparticle state $\ket{P,J,h}$, where $h$ denotes the helicity.
The coefficients $\mathcal C_{\mathcal I \to *}^{J,h}$ can be seen as elements of the complex vector space $V_{\mathcal I \to *}$, \textit{i.e.}, $\ket{\mathcal C_{\mathcal I \to *}^{J,h}}\in V_{\mathcal I \to *}$, and, correspondingly, the angular momentum basis elements are given by
\begin{equation}
    \ket{\mathcal B^{J}_{i\to f}} = \sum_h \ket{\mathcal C_{i \to *}^{J,h}} \otimes \ket{\mathcal C_{* \to f}^{J,h}} \in V_{i\to f}\,,
\end{equation}
where
$\ket{\mathcal C^{J,h}_{* \to \mathcal I}}=\ket{(\mathcal C^{J,h}_{\mathcal I \to *})^*} \in V_{\ast \to \mathcal I}$ and $V_{i\to f} = V_{i \to *}\otimes V_{* \to f}$.
The partial wave coefficients $a_{i\to f}^J$ can then be conveniently obtained by extending the inner product 
in $V_{\mathcal I \to *}$~\cite{Shu:2021qlr} to the
vector space $V_{i\to f}$ as follows
\begin{align}
    a_{i\to f}^J &= \frac{1}{2J+1}\langle \mathcal B^J_{i\to f}|\mathcal A_{i\to f}\rangle \nonumber \\
    &=
    \frac{1}{2J+1}\int \text{d}\Phi_i\, \text{d}\Phi_f\, \mathcal A_{i\to f}\left(\mathcal B^J_{i\to f}\right)^*\,,
    \label{eq:partial_wave}
\end{align}
where $\text{d}\Phi_{\mathcal I}$ refers to the Lorentz invariant phase space measure associated with $\ket{\mathcal I}$~\footnote{Explicit parameterizations for the $n$-body phase space in terms of spinor-helicity variables are available in \cite{Zwiebel:2011bx,Mastrolia:2009dr,Cox:2018wce,Larkoski:2020thc,EliasMiro:2020tdv}.}.
Notice that we crucially chose the basis normalization
$\langle \mathcal B^J_{i\to f}|\mathcal B^{J'}_{i\to f}\rangle = (2J+1)\delta^{JJ'}$
in order to obtain the partial wave unitarity bounds in the standard form.
Indeed, this choice implies
\begin{equation}
    \int \text d \Phi_X\,\ket{\mathcal B^{J}_{i\to X}}\otimes \ket{\mathcal B_{X\to f}^{J'}} = \ket{\mathcal B_{i\to f}^J} \delta^{JJ'}\,,
\end{equation}
which, when applied to the generalized optical theorem
\begin{equation}
    \ket{\mathcal A_{i\to f}} - \ket{\mathcal A_{f\to i}^*} = i \sum_X \int \text d \Phi_X\, \ket{\mathcal A_{i\to X}}\otimes\ket{\mathcal A_{f\to X}^* }
\end{equation}
leads to  
\begin{equation}
    a_{i\to f}^J - \left(a_{f\to i}^J\right)^* = i \sum_X a_{i\to X}^{J} \left(a_{f\to X}^{J}\right)^*
\end{equation}
and, in turn, to the partial wave unitarity bounds
\begin{gather}
    \left|\Re a_{i\to i}^J \right| \le 1 \,,~~~~  0 \le \Im a_{i\to i}^J \le 2\,,~~~~
    \left|a^J_{i\to f}\right| \le 1 \,,
\end{gather}
where $i\neq f$.
In practice, the determination of the angular momentum basis elements $\ket{\mathcal B^{J}_{i\to f}}$, required by the partial wave decomposition,
can be achieved without passing through the Poincaré Clebsch-Gordan coefficients $\mathcal C_{\mathcal I \to *}^{J,h}$, as one can proceed as follows:
\begin{enumerate}
    \item Find a set of kinematic monomials in spinor-helicity variables that are consistent with the particle helicities, span $V_{i\to f}$, and are independent after accounting for momentum conservation and Schouten identities~\footnote{Public Mathematica packages that allow one to perform this task are available in~\cite{DeAngelis:2022qco,Li:2022tec}.}.
    \item Apply the Pauli-Lubanski operator squared~\cite{Witten:2003nn}
    \begin{align}
    \mathbb W^2_{\mathcal I} &= \frac{1}{8}\mathbb P_{\mathcal I}^2 \left(\epsilon^{\alpha\gamma}\epsilon^{\beta\delta}\mathbb M_{\mathcal I,\alpha\beta}\mathbb M_{\mathcal I,\gamma\delta}+\epsilon^{\dot\alpha\dot\gamma}\epsilon^{\dot\beta\dot\delta}\widetilde{\mathbb M}_{\mathcal I,\dot\alpha\dot\beta}\widetilde{\mathbb M}_{\mathcal I,\dot\gamma\dot\delta}\right)\nonumber \\
    &\quad 
    +\frac{1}{4}\mathbb P_{\mathcal I}^{\alpha\dot\alpha}\mathbb P_{\mathcal I}^{\beta\dot\beta}\mathbb M_{\mathcal I,\alpha\beta}\widetilde{\mathbb M}_{\mathcal I,\dot\alpha\dot\beta}\,,
\end{align}
where
\begin{align}
    \mathbb P_{\mathcal I}^{\alpha\dot\alpha} &= \sum_{i\in \mathcal I}\lambda_i^\alpha \widetilde \lambda_i^{\dot\alpha}\,,\\
    \mathbb M_{\mathcal I}^{\alpha\beta} &= \sum_{i\in \mathcal I}\left(\lambda_{i}^\alpha\frac{\partial}{\partial \lambda_{i,\beta}}+\lambda_{i}^\beta \frac{\partial}{\partial \lambda_{i,\alpha}}\right)\,,\\
    \widetilde{\mathbb M}_{\mathcal I}^{\dot\alpha\dot\beta} &= \sum_{i\in \mathcal I}\left(\widetilde\lambda_{i}^{\dot\alpha}\frac{\partial}{\partial \widetilde\lambda_{i,\dot\beta}}+\widetilde\lambda_{i}^{\dot\beta}\frac{\partial}{\partial \widetilde\lambda_{i,\dot\alpha}}\right)\,,
\end{align}
and $\mathcal I=i,f$~\footnote{We use the convention $\epsilon^{12}=\epsilon^{\dot 1 \dot 2} =  1$, where spinor indices are raised and lowered as $\lambda ^\alpha = \epsilon^{\alpha\beta}\lambda_\beta$ and $\tilde\lambda_{\dot \alpha} = \epsilon_{\dot\alpha \dot\beta}\tilde\lambda^{\dot\beta}$. Angle and square brackets are then defined as $\agl{i}{j} = \lambda_i^\alpha \lambda_{j,\alpha}$ and $\sqr{i}{j} = \tilde\lambda_{i,\dot\alpha}\tilde\lambda_j^{\dot\alpha}$, respectively.}, to these monomials and construct the associated matrix.
\item Find the eigenvectors of the above matrix and the associated values of angular momentum from the Casimir eigenvalues $-P^2_{\mathcal I}J_{\mathcal I}(J_{\mathcal I}+1)$.
\item Normalize the elements of the so obtained orthogonal basis so that their norm is equal to $\sqrt{2J_{\mathcal I}+1}$.
\end{enumerate}

As expected, for $2\to 2$ scattering, the basis $\ket{\mathcal B^{J}_{i\to f}}$ is proportional to the Wigner $d$-matrix~\cite{Jiang:2020rwz}. 
Moreover, we have explicitly verified that the
partial wave coefficients for $N\to M$ scattering (with $N,M \geq 2$) exactly reproduce the elements of the reduced scattering matrix~\cite{Chang:2019vez,Falkowski:2019tft,Cohen:2021ucp,Mahmud:2025wye}, which captures only the $s$-wave contribution.

As an interesting result of our analysis, in the Appendix we report the angular momentum basis for $2 \to 3$ amplitudes which is valid for generic values of $J$.

\medskip
\paragraph{EFT of gravity and light-by-light scattering.}
As a first application of the above method, we analyze the unitarity bounds for the EFT of gravity and light-by-light scattering involving operators with four gravitons and photons, respectively.
The versatility of the spinor-helicity formalism in handling particles of arbitrary spin~\cite{Arkani-Hamed:2017jhn} allows us to approach the problem in a unified manner.

We focus on the P- and CP-violating Lagrangian:
\begin{equation}
\label{eq:Lag-quartic}
    \frac{\mathcal L^{(S)}}{\sqrt{-g}} = c_1^{(S)}( \mathcal Q^{(S)})^2 + c_2^{(S)} (\widetilde{\mathcal Q}^{(S)})^2 + c_3^{(S)}  \mathcal Q^{(S)} \widetilde{\mathcal Q}^{(S)}
\end{equation}
where
\begin{align}
    \mathcal Q^{(1)} &= F_{\mu\nu}F^{\mu\nu},
    &
    \widetilde{\mathcal Q}^{(1)} &= F_{\mu\nu}\widetilde F^{\mu\nu},\\
    \mathcal Q^{(2)} &= \Mpl^2 R_{\mu\nu\rho\sigma}R^{\mu\nu\rho\sigma},
    &
    \widetilde{\mathcal Q}^{(2)} &= \Mpl^2 R_{\mu\nu\rho\sigma}\widetilde R^{\mu\nu\rho\sigma},
\end{align}
$\widetilde F_{\mu\nu}=\frac{1}{2} \epsilon_{\mu\nu\rho\sigma}F^{\rho\sigma}$ is the dual photon field-strength tensor, $\widetilde R_{\mu\nu\rho\sigma} = \frac{1}{2} \epsilon_{\mu\nu\alpha\beta}R^{\alpha\beta}{}_{\rho\sigma}$ is the dual Riemann tensor~\footnote{The convention for the Levi-Civita tensor is $\epsilon^{0123}=1/\sqrt{-g}$.}, $\Mpl^2=(8\pi G_{\text{N}})^{-1}$ is the Planck mass squared,
and $c_i^{(S)}$ are dimensionful Wilson coefficients with $[c_i^{(S)}]=-4S$. 
For $S=1$, we recover the electromagnetic Euler-Heisenberg Lagrangian~\cite{Heisenberg:1936nmg}, while, for $S=2$, we obtain the eight-derivative corrections to the gravitational Einstein-Hilbert action~\cite{Endlich:2017tqa}.
In both cases, the three operators appearing in Eq.~\eqref{eq:Lag-quartic} constitute 
a basis for such effective dimension-$(4S+4)$ operators~\cite{Ruhdorfer:2019qmk,Li:2023wdz}.

Denoting by $i=\{(1^{+S},2^{+S})$, $ (1^{+S},2^{-S})$, $ (1^{-S},2^{-S})\}$ all the possible initial two-particle states and by $f=\{(3^{+S},4^{+S})$, $ (3^{+S},4^{-S})$, $(3^{-S},4^{-S})\}$ the final ones, the full four-particle scattering matrix reads
\begin{widetext}
\begin{equation}\label{eq:amplitude}
    \ket{\mathcal A_{i\to f}} = 
    \begin{pmatrix}
        \substack{8c_+^{(S)}\agl{1}{2}^{2S}\sqr{3}{4}^{2S}} & 0 
        & \substack{8c_-^{(S)}(\agl{1}{2}^{2S}\agl{3}{4}^{2S}\\+\agl{1}{3}^{2S}\agl{2}{4}^{2S}+\agl{1}{4}^{2S}\agl{2}{3}^{2S})}
        \\
        0 & \substack{8c_+^{(S)}\agl{1}{4}^{2S}\sqr{2}{3}^{2S}} 
        & 0
        \\
        \substack{8(c_-^{(S)})^*(\sqr{1}{2}^{2S}\sqr{3}{4}^{2S}\\+\sqr{1}{3}^{2S}\sqr{2}{4}^{2S}+\sqr{1}{4}^{2S}\sqr{2}{3}^{2S})} & 0 
        & \substack{8c_+^{(S)}\agl{3}{4}^{2S}\sqr{1}{2}^{2S}}
    \end{pmatrix}
\end{equation}
\end{widetext}
with $c_+^{(S)}= c_1^{(S)} + c_2^{(S)}\in \mathbb{R}$ and $c_-^{(S)}= c_1^{(S)} - c_2^{(S)} + i c_3^{(S)}\in \mathbb{C}$.
A simple way to obtain it relies on writing the photon field strength in the spinor space as $F_{\alpha \dot \alpha \beta \dot \beta} \sim \epsilon_{\alpha\beta}\tilde\lambda_{\dot \alpha}\tilde\lambda_{\dot \beta}+\epsilon_{\dot\alpha\dot\beta}\lambda_\alpha \lambda_\beta$ and similarly for the Riemann tensor $R_{\alpha\dot\alpha\beta\dot\beta\gamma\dot\gamma\delta\dot\delta}\sim \epsilon_{\alpha\beta}\epsilon_{\gamma\delta}\tilde\lambda_{\dot\alpha}\tilde\lambda_{\dot\beta}\tilde\lambda_{\dot\gamma}\tilde\lambda_{\dot\delta} + \epsilon_{\dot\alpha\dot\beta}\epsilon_{\dot\gamma\dot\delta}\lambda_{\alpha}\lambda_{\beta}\lambda_{\gamma}\lambda_{\delta}$~\cite{Arkani-Hamed:2020blm}.

The kinematic vector space for the maximally helicity violating amplitude $\mathcal A_{1^{-S},2^{-S}\to 3^{+S},4^{+S}}$ in Eq.~\eqref{eq:amplitude}, has dimension $2S+1$ and is spanned by $\{\ket{m_k}\}_{k=1}^{2S+1}$, with
\begin{equation}
    \ket{m_k} = \sqr{1}{2}^{2S-k+1}\sqr{1}{4}^{k-1}\sqr{2}{3}^{k-1}\sqr{3}{4}^{2S-k+1} \,,
\end{equation}
which constitutes a basis.
Of course, the same is true for $\mathcal A_{1^{+S},2^{+S}\to 3^{-S},4^{-S}}$, provided that all the square inner products are substituted with angle ones.
In this basis, the Pauli-Lubanski operator squared is represented by an upper bidiagonal matrix:
\begin{equation}
    \mathbb W^2_{12}=-s 
    \begin{pmatrix}
        0 & 1 & 0 & \cdots & \cdots & \cdots & 0 \\
        0 & 2 & 4 & \ddots & \ddots & \ddots & \vdots \\
        0 & 0 & 6 & 9 & \ddots & \ddots & \vdots \\
        \vdots & \ddots & 0 & 12 & 16 &\ddots & \vdots \\
        \vdots & \ddots & \ddots & 0 & 20 & \ddots & 0 \\
        \vdots & \ddots & \ddots & \ddots & \ddots & \ddots & 4S^2 \\
        0 & \cdots & \cdots & \cdots & 0 & 0 & 2S(2S+1) 
    \end{pmatrix}
\end{equation}
where $s=(p_1+p_2)^2$.
In fact, its action on the basis monomials $\ket{m_k}$, with $k>1$, is
\begin{multline}
    \!\!\!\!\!\!\mathbb W^2_{12} \ket{m_k} = -s
    \frac{k-1}{2}\sqr{1}{2}^{2S-k+1}\sqr{1}{4}^{k-2}\sqr{2}{3}^{k-2}\sqr{3}{4}^{2S-k+1} \\
     \times 
    \big((k+1)\sqr{1}{4}\sqr{2}{3}+(k-1)(\sqr{1}{3}\sqr{2}{4}+\sqr{1}{2}\sqr{3}{4})\big)\,,
\end{multline}
which, upon using the Schouten identity $\sqr{1}{3}\sqr{2}{4}=\sqr{1}{2}\sqr{3}{4}+\sqr{1}{4}\sqr{2}{3}$, becomes
\begin{equation}
\!\!    \mathbb W^2_{12} \ket{m_k} = -s\left((k-1)^2\ket{m_{k-1}}+k(k-1)\ket{m_k}\right)\,.
\end{equation}
The eigenvalues of this matrix are given by its diagonal entries and correspond to $J_{12}=0,1,\dotsc,2S$. 

Instead, for all the amplitudes in Eq.~\eqref{eq:amplitude} that are proportional to $c_+^{(S)}$, the dimension of the kinematic vector space is $1$, and thus such amplitudes have a definite angular momentum.
In particular, $\mathcal A_{1^{\pm S},2^{\pm S}\to 3^{\pm S},4^{\pm S}}$ has $J_{12}=0$, while $\mathcal A_{1^{+ S},2^{- S}\to 3^{+ S},4^{- S}}$ has $J_{12}=2S$, since 
\begin{equation}
    \mathbb  W^2_{12}\agl{1}{2}^{2S}\sqr{3}{4}^{2S} = \mathbb  W^2_{12}\agl{3}{4}^{2S}\sqr{1}{2}^{2S}=0
\end{equation}
and 
\begin{align}
    \mathbb  W^2_{12}\agl{1}{4}^{2S}\sqr{2}{3}^{2S} = -s[2S(2S+1)]\agl{1}{4}^{2S}\sqr{2}{3}^{2S}\,.
\end{align}

\begin{figure*}[h!t!b!]
\centering
\includegraphics[width=0.9\textwidth]{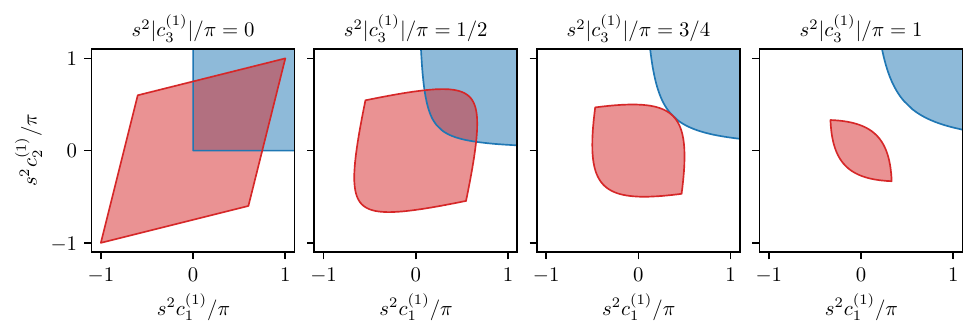}
\includegraphics[width=0.9\textwidth]{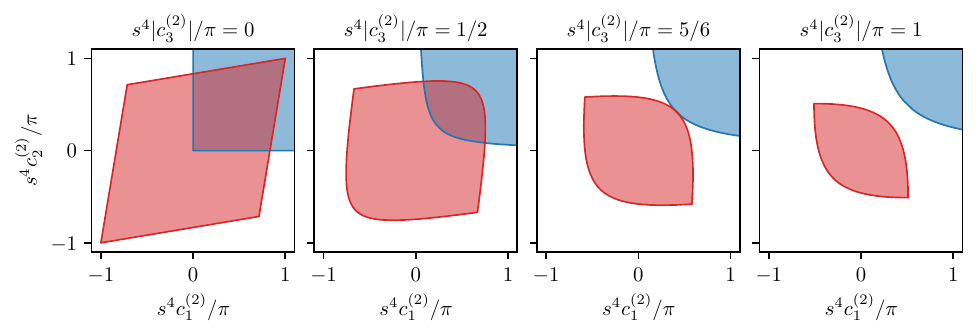}
\caption{
Allowed regions for the Euler-Heisenberg EFT (upper plot) and the
EFT of gravity (lower plot) by partial wave unitarity bounds (red) 
and positivity bounds (blue) in the $s^{2S}c_{1}^{(S)}/\pi$ and $s^{2S}c_{2}^{(S)}/\pi$ plane for different values of $c_3^{(S)}$,
see Eq.~\eqref{eq:Lag-quartic}.
\label{fig:s1}}
\end{figure*}

The normalized eigenvectors corresponding to $J_{12}=0$ are then
\begin{equation}
    \ket{\mathcal B_{i\to f}^0} 
    =
    \frac{16\pi}{s^{2S}}
    \begin{pmatrix}
        \agl{1}{2}^{2S}\sqr{3}{4}^{2S} & 0 
        & \agl{1}{2}^{2S}\agl{3}{4}^{2S}\\
        0 & 0 
        & 0\\
        \sqr{1}{2}^{2S}\sqr{3}{4}^{2S} & 0 
        & \sqr{1}{2}^{2S}\agl{3}{4}^{2S}
        \end{pmatrix}\,,
\end{equation}
and, by projecting the amplitudes in Eq.~\eqref{eq:amplitude} onto them, it follows that the corresponding partial waves read~\footnote{
In this case, the phase space integral can be achieved, \textit{e.g.}, by parameterizing the final-state spinors as~\cite{Mastrolia:2009dr,Bresciani:2024shu}
\begin{equation*}
    \begin{pmatrix}
        \lambda_3 \\ \lambda_4
    \end{pmatrix}
    =
    \frac{1}{\sqrt{1+z\bar z}}
    \begin{pmatrix}
        1 & \bar z \\
        -z & 1
    \end{pmatrix}
    \begin{pmatrix}
        \lambda_1 \\ \lambda_2
    \end{pmatrix}
\end{equation*}
and writing the phase space measure as 
\begin{equation*}
    \int \text{d}\Phi_{34} =-\frac{1}{2} \frac{1}{8\pi}\oint \frac{\text{d}z}{2\pi i}\int \frac{\text{d}\bar z}{(1+z\bar z)^2}\,,
\end{equation*}
where we included the $1/2$ symmetry factor due to indistinguishable particles.
This yields the factors
\begin{equation*}
    -\Res_{z=0}\int \text{d}\bar z\frac{1+(z\bar z)^{2S}+(1+z\bar z)^{2S}}{(1+z\bar z)^{2S+2}} = \frac{2S+3}{2S+1}
\end{equation*}
that appear in Eq.~\eqref{eq:partialwavecoeffs}.}
\begin{equation}\label{eq:partialwavecoeffs}
    a^{0}_{i\to f} =
    \frac{s^{2S}}{2\pi}
    \begin{pmatrix}
        c_+^{(S)} & 0 
        & \frac{2S+3}{2S+1}c_-^{(S)} \\
        0 & 0 
        & 0\\
        \frac{2S+3}{2S+1}(c_-^{(S)})^* & 0 
        & c_+^{(S)}
    \end{pmatrix}\,,
\end{equation}
whose non-vanishing eigenvalues are
\begin{equation}
    \frac{s^{2S}}{2\pi}\left(c_+^{(S)}\pm \frac{2S+3}{2S+1}|c_-^{(S)}| \right)\,.
\end{equation}
As a result, we find that the strongest unitarity bound is provided by the partial wave with $J_{12}=0$:
\begin{equation}
\label{eq:PUbounds}
    \mathscr{U}:\quad \frac{s^{2S}}{2\pi}\left(|c_+^{(S)}|+\frac{2S+3}{2S+1}|c_-^{(S)}|\right) \le 1\,.
\end{equation}
The regions of the parameter space satisfying the partial wave unitarity condition of Eq.~\eqref{eq:PUbounds} are shown in red in Fig.~\ref{fig:s1} for the cases of $S=1,2$, respectively \footnote{In deriving these bounds for the gravitational EFT ($S=2$), we explicitly neglected the universal long-range contribution from single-graviton exchange
\begin{equation*}
    \ket{\mathcal A_{1^{+2},2^{+2} \to 3^{+2},4^{+2}}} = \frac{1}{\Mpl^2}\frac{\agl{1}{2}^4 \sqr{3}{4}^4}{stu}\,,
\end{equation*}
responsible for the well-known forward and backward singularities that induce infrared divergences in the partial wave projections of the scattering amplitude. Such divergences are a well-understood feature of long-range interactions and can be systematically handled by factoring out a universal infrared-divergent phase \cite{Weinberg:1965nx} common to all partial waves. This universal phase, which arises from the resummation of iterated single-graviton exchange (the gravitational Born series), does not affect physical observables and can be stripped from the $S$-matrix, as done for example in \cite{Blas:2020dyg,Bellazzini:2025bay}, yielding infrared-finite amplitudes suitable for unitarity and positivity analyses, and leaving the bounds presented here robust. The resolution scale $\mathcal E$ introduced by this procedure \cite{Bellazzini:2025bay} admits a physical interpretation as the energy resolution of a macroscopic detector for soft gravitons. 
Alternative approaches to deal with the above singularity include the smearing method of Ref.~\cite{Caron-Huot:2021rmr} and the dimensional reduction technique of Ref.~\cite{Bellazzini:2019xts}.}.

Complementary bounds on our EFT parameter space can be obtained by imposing positivity bounds, stemming from the analyticity and causality of scattering amplitudes~\cite{Adams:2006sv}. In particular, they impose specific inequalities on the Wilson coefficients to ensure compatibility with a well-behaved UV completion.
In our scenarios, they read $|c_-^{(S)}| \le c_+^{(S)}$, or, equivalently,
\begin{align}
\label{eq:positivitybounds}
    \mathscr{P}: \quad 
    c_{1,2}^{(S)} \ge 0 \,\,\,\,\,\text{and}\,\,\,\,\,
    (c_3^{(S)})^2 \le 4\, c_{1}^{(S)} c_{2}^{(S)}\,,
\end{align}
and are valid for both $S=1$~\cite{Adams:2006sv,Rebhan:2017zdx,Remmen:2019cyz,Henriksson:2021ymi,Arkani-Hamed:2020blm} 
and $S=2$~\cite{Gruzinov:2006ie,Bellazzini:2015cra,Endlich:2017tqa,Arkani-Hamed:2020blm,Caron-Huot:2022ugt,Bellazzini:2022wzv}.
The blue regions in Fig.~\ref{fig:s1} satisfy positivity bounds. Interestingly, we emphasize that the viable parameter space is significantly reduced once both positivity and partial wave unitarity bounds are imposed. 

To highlight the impact of positivity bounds, we compare the volume of the parameter space allowed by unitarity, $\text{Vol}(\mathscr U)$, with the one obtained after imposing the positivity constraints, $\text{Vol}(\mathscr U \cap \mathscr P)$.
In particular, for fixed values of the center-of-mass energy, we find
\begin{equation}
\frac{\text{Vol}(\mathscr U \!\cap\! \mathscr P)}{\text{Vol}(\mathscr U)} = \frac{1}{32}\!\left(\frac{2S+3}{S+1}\right)^{\!2}\!
    \approx \!
    \begin{cases}
        0.20 & \!\text{for }S\!=\!1\,,\\
        0.17 & \!\text{for }S\!=\!2\,,
    \end{cases}
\end{equation}
which is a monotonically decreasing function of $S$.

\medskip
\paragraph{$N\to M$ scattering amplitudes.}
Another important advantage of our method over standard techniques 
is its capability of addressing unitarity 
bounds for $N \to M$ processes (with $N,M \geq 2$).
As an illustrative example, we consider the dimension-six 
SMEFT interactions~\cite{Grzadkowski:2010es}
\begin{align}
    \mathcal L^{(6)} &\supset 
    C_{eH}^{pr}\left (H^\dagger H\right)\overline \ell_p e_r H + 
    C_{dH}^{pr}\left (H^\dagger H\right)\overline q_p d_r H  
    \nonumber\\ & + C_{uH}^{pr}\left(H^\dagger H\right)\overline q_p u_r \widetilde H + \text{h.c.}
    \label{eq:SMEFT_6}
\end{align}
where $\widetilde H = i \sigma^2 H^*$ and $p,r$ are flavour indices.
As is well known, $\mathcal L^{(6)}$ induces modifications to the fermionic Yukawa couplings, which are a target of the High-Luminosity LHC program,
and generates new contributions to multi-boson interactions, which might be probed at future colliders.

To obtain the unitarity bounds, we performed a coupled channel analysis involving all possible contact amplitudes for
$\phi_a\phi_b \to \phi_c \psi_i^\pm\overline \psi_j^\pm$ and $\psi_i^\pm \overline\psi_j^\pm\to \phi_a\phi_b\phi_c$, where $\{\phi_a\}_{a=1}^4$ denote the real components of $H=(\phi_1+i\phi_3,\phi_2+i\phi_4)^T/\sqrt{2}$ and $\psi_i,\psi_j$ any Dirac fermions.
We find that the strongest unitarity bound arises from the $J=0$ partial waves and reads
\begin{equation}
\!\sqrt{\Tr[3C_{uH}C_{uH}^\dagger + 3C_{dH}C_{dH}^\dagger + C_{eH}C_{eH}^\dagger]}  \le \frac{ 32\pi^2}{\sqrt{3}s}\,,\label{eq:SMEFTdim6UB}
\end{equation}
where the trace is understood over flavour indices.
As for the $\phi_a\phi_b \to \phi_c \overline \psi_i^+ \psi_j^+$ amplitude
\begin{equation}
    \ket{\mathcal A_{1_{\phi_1},2_{\phi_1}\to 3_{\phi_1},4^+_{\overline \nu_p},5^+_{e_r}}} =
    \frac{3\sqrt{2}}{2}(C_{eH}^{pr})^* \sqr{5}{4}\,,
\end{equation}
the relevant vector basis element of the Appendix is
\begin{align}
    \ket{\mathcal B^0_{1^0,2^0 \to 3^0,4^{\frac{1}{2}},5^{\frac{1}{2}}}} &= \frac{32\sqrt{6}\pi^2}{s}\sqr{5}{4}\,,
\end{align}
where $s=(p_1+p_2)^2$. Therefore, without performing any phase space integral,
we get the $J=0$ partial wave
\begin{equation}
    a^0_{1_{\phi_1},2_{\phi_1}\to 3_{\phi_1},4^+_{\overline \nu_p},5^+_{e_r}} = \frac{\sqrt{3}s}{64\sqrt{2}\pi^2}(C_{eH}^{pr})^*\,,
\end{equation}
where we included a $1/\sqrt{2}$ factor accounting for the identical particles in the initial state.

Instead, the unitarity bound from $2 \to 2$ amplitudes, arising from Eq.~(\ref{eq:SMEFT_6}) picking one VEV $v$ out of a Higgs doublet, reads
\begin{equation}
\!\! \sqrt{\Tr[3C_{uH}C_{uH}^\dagger \!+ 3C_{dH}C_{dH}^\dagger \!+ C_{eH}C_{eH}^\dagger]}
    \!\le \frac{ 4\sqrt{2}\pi }{\sqrt{3 s v^2}}\,,
\end{equation}
and is weaker than that of Eq.~\eqref{eq:SMEFTdim6UB} for $\sqrt{s} > 4 \pi v\sqrt{2}$.

Next, we consider the following dimension-eight SMEFT interactions
\begin{align}
    \mathcal L^{(8)} &\supset C_{X^3H^2}\left(H^\dagger H\right)f^{ABC}X^A_{\mu}{}^{\nu} X^B_{\nu}{}^{\rho} X^C_{\rho}{}^{\mu} \nonumber \\
    &+ \widetilde C_{X^3H^2}\left(H^\dagger H\right)f^{ABC}X^A_{\mu}{}^{\nu} X^B_{\nu}{}^{\rho} \widetilde X^C_{\rho}{}^{\mu}\,,
    \label{eq:SMEFT_8}
\end{align}
where $X^A_{\mu\nu}$ is the field strength associated with a generic non-Abelian gauge group $G$.
The strongest unitarity constraint stems from the $J=0$ partial waves and reads
\begin{equation}
    \sqrt{C_{X^3H^2}^2+\widetilde C_{X^3H^2}^2} \le
    \frac{32 \sqrt{10} \pi^2}{s^2}\frac{1}{\sqrt{C_2(G) d(G)}}\,, 
    \label{eq:dim8unbroken}
\end{equation}
where $C_2(G)$ is the quadratic Casimir of the adjoint representation of $G$ and $d(G)$ is its dimension~\footnote{$C_2(G)$ is defined by $f^{ACD}f^{BCD} = C_2(G)\delta^{AB}$. In particular, the structure constants of the $SU(N)$ group, which has dimension 
$d(SU(N)) = N^2 -1$, are normalized such that $C_2(SU(N)) = N$.}.
This bound is obtained via a coupled channel analysis which includes all the processes $\phi_a \phi_b \to X^A{}^\pm X^B{}^\pm X^C{}^\pm$. 
Taking the related amplitudes 
\begin{align}
    \ket{\mathcal A_{1_{\phi_a},2_{\phi_b} \to 3_{X^A}^-,4_{X^B}^-,5_{X^C}^-}} &= 
    3\sqrt{2}(iC_{X^3H^2}-\widetilde C_{X^3H^2})\delta_{ab} \nonumber \\
    &\quad \times f^{ABC} \agl{3}{4}\agl{4}{5}\agl{3}{5}\,, \\
    \ket{\mathcal A_{1_{\phi_a},2_{\phi_b} \to 3_{X^A}^+,4_{X^B}^+,5_{X^C}^+}} &= 
    3\sqrt{2}(iC_{X^3H^2}+\widetilde C_{X^3H^2})\delta_{ab} \nonumber \\
    &\quad \times f^{ABC} \sqr{4}{3}\sqr{5}{4}\sqr{5}{3}\,,
\end{align}
and the relevant vector basis element of the Appendix
\begin{equation}
    \ket{\mathcal B^0_{1^0,2^0 \to 3^1,4^{1},5^{1}}} = \frac{64\sqrt{30}\pi^2}{s^2}\sqr{4}{3}\sqr{5}{3}\sqr{5}{4}\,,
\end{equation}
we can immediately obtain the partial wave coefficients by including the $1/\sqrt{2}$ symmetry factor:
\begin{align}
    a^0_{1_{\phi_a},2_{\phi_b} \to 
    3_{\!X^{\!A}}^\pm,4_{\!X^{\!B}}^\pm,5_{\!X^{\!C}}^\pm} \!\!=\!
    \frac{3 s^2 \delta_{ab}f^{ABC}}{64\sqrt{30}\pi^2}(iC_{X^3H^2}\pm \widetilde C_{X^3H^2})
    \,.
\end{align}

\begin{figure}[h!t!b!]
\centering
\includegraphics[width=\columnwidth]{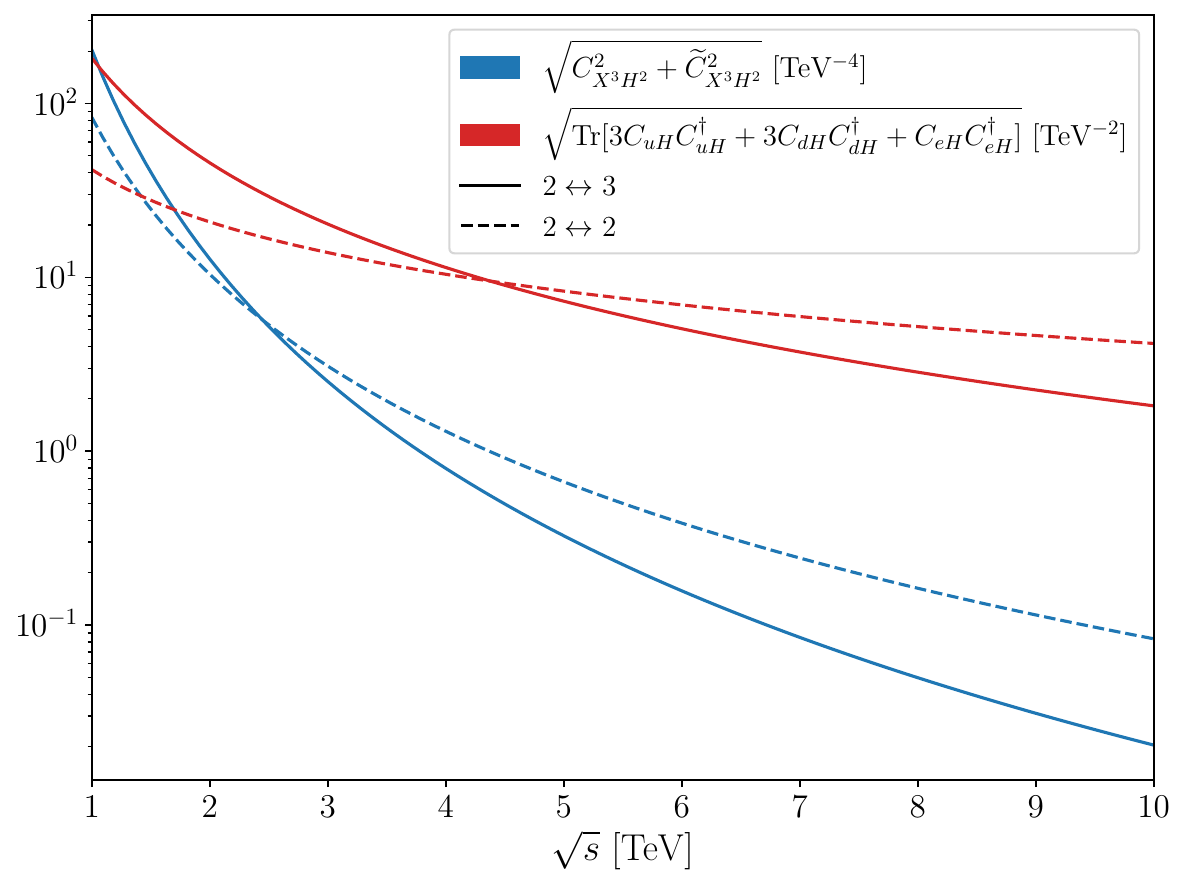}
\caption{
Partial wave unitarity bounds on the Wilson coefficients of SMEFT operators, see Eqs.~\eqref{eq:SMEFT_6} and \eqref{eq:SMEFT_8},
\textit{vs.}~$\sqrt{s}$ 
from $2\to 2$ processes (dashed lines) and $2\to 3$ processes (solid lines).
$C_{X^3H^2}$ and $\widetilde C_{X^3H^2}$ refer to $G=SU(3)$.}
\label{SMEFT_plot}
\end{figure}

Instead, $h X^A{}^\mp \to X^B{}^\pm X^C{}^\pm$ (corresponding to $J=1$) are the only available $2\to 2$ channels and the related unitarity bound is
\begin{equation}
    \sqrt{C_{X^3H^2}^2+\widetilde C_{X^3H^2}^2} \le \frac{8\sqrt{2}\pi}{v s^{3/2}}\frac{1}{\sqrt{C_2(G)}}\,,
\end{equation}
which is weaker than that of Eq.~\eqref{eq:dim8unbroken} for $\sqrt{s} > 4\pi v\sqrt{5/d(G)}$.
In Fig.~\ref{SMEFT_plot}, we show the unitarity bounds 
arising from $2\to 2$ and $2\to 3$ processes as induced 
by $\mathcal L^{(6)}$ and $\mathcal L^{(8)}$, see 
Eqs.~(\ref{eq:SMEFT_6}) and (\ref{eq:SMEFT_8}).

\medskip
\paragraph{Conclusions.}
In this Letter, we elaborated on the application of on-shell methods for generalizing the formulation of partial wave unitarity bounds. 

In particular, we provided the general angular momentum basis for $2 \to 3$ amplitudes, extending previous results obtained by the reduced scattering matrix method in the limit of $J=0$.
As an illustration of our results, we discussed unitarity bounds for $2 \to 3$ scattering processes as induced by some dimension-six and -eight SMEFT operators, see Fig.~\ref{SMEFT_plot}, that might be probed at high-energy future colliders.
Moreover, we discussed for the first time unitarity bounds for spin-2 or higher-spin theories---relevant for effective field theories of gravity---that are not approachable by standard methods, see Fig.~\ref{fig:s1}. 

Finally, we emphasized the complementarity of positivity and partial wave unitarity bounds to probe the parameter space of effective field theories, 
see Fig.~\ref{fig:s1}.

\medskip
\paragraph{Acknowledgments.}
This work received funding by the INFN Iniziative Specifiche AMPLITUDES and APINE and from the European Union’s Horizon 2020 research and innovation programme under the Marie Sklodowska-Curie grant agreements n.~860881 – HIDDeN, n.~101086085 – ASYMMETRY. 
This work was also partially supported by the Italian MUR Departments of Excellence grant 2023-2027 “Quantum Frontiers”.
The work of P.P.~is supported by the European Union – Next Generation EU and by the Italian Ministry of University and Research (MUR) via the PRIN 2022 project n.~2022K4B58X – AxionOrigins. 
G.L.~gratefully acknowledges financial support from the Swiss National Science Foundation (Project No.\ TMCG-2\_213690).

\bibliography{letter_bibliography}

\clearpage

\onecolumngrid
\appendix*

\section{Appendix: Angular momentum basis for \texorpdfstring{$2 \to 3$}{2-to-3} amplitudes}

\twocolumngrid
Here, we list the normalized kinematic polynomials with definite angular momentum $J_{12}$ relevant for $2 \to 3$ helicity amplitudes.
They are denoted by $\ket{\mathcal B^{J_{12}}_{i\to f}}$, where $i=(1^{-h_1},2^{-h_2})$ and $f=(3^{h_3},4^{h_4},5^{h_5})$, while $s_{12} = (p_1+p_2)^2$.
We report only those with $\sum_{i=1}^5 h_i \ge 0$. The helicity configurations not listed here can be recovered from the listed ones
by applying a parity flip, \textit{i.e.}~applying the following substitutions: $h_i \to -h_i$ and $\agl{i}{j}\leftrightarrow \sqr{j}{i}$ for all $i,j$.
Notice that in many cases the reported subspace of $V_{i\to f}$ with fixed angular momentum is degenerate.
This degeneracy is due to the fact that a subset of particles in the final state can have more than one angular momentum value.

{\footnotesize
\begin{eqnarray*}
    \toprule
    \toprule
    (h_1,h_2;h_3,h_4,h_5) &
    \,\,\,\,\, 
    J_{12}
    \,\,\,\,\, 
    & \ket{\mathcal B^{J_{12}}_{i \to f}}\\ 
    \midrule
    \midrule
    (-1,-1;0,1,1) & 0 & 64\sqrt{3}\pi^2s_{12}^{-5/2} \agl{1}{2}^2 \sqr{5}{4}^2 \\
    (-1,-1;\hf,\hf,1) & 0 & 64\sqrt{6}\pi^2s_{12}^{-5/2} \agl{1}{2}^2 \sqr{5}{3}\sqr{5}{4} \\
    (-1,-1;1,1,1) & 0 & 64\sqrt{30}\pi^2s_{12}^{-3} \agl{1}{2}^2 \sqr{4}{3}\sqr{5}{3} \sqr{5}{4} \\
    \midrule
    (-1,-\hf;-\hf,1,1) & \hf & 64\sqrt{30}\pi^2s_{12}^{-5/2} \agl{1}{2}\agl{1}{3}\sqr{5}{4}^2 \\
    (-1,-\hf;0,\hf,1) & \hf & 64\sqrt{15}\pi^2s_{12}^{-5/2} \agl{1}{2}\agl{1}{4}\sqr{5}{4}^2 \\
    && 128\sqrt{10}\pi^2s_{12}^{-5/2} \agl{1}{2}\agl{1}{3}\sqr{5}{3}\sqr{5}{4} \\
    (-1,-\hf;\hf,\hf,\hf) & \hf & 64\sqrt{30}\pi^2s_{12}^{-5/2} \agl{1}{2}\agl{1}{3}\sqr{4}{3}\sqr{5}{3} \\
    && 64\sqrt{30}\pi^2s_{12}^{-5/2} \agl{1}{2}\agl{1}{4}\sqr{4}{3}\sqr{5}{4} \\
    && 64\sqrt{30}\pi^2s_{12}^{-5/2} \agl{1}{2}\agl{1}{5}\sqr{5}{3}\sqr{5}{4} \\
    (-1,-\hf;\hf,1,1) & \hf & 384\sqrt{2}\pi^2s_{12}^{-3} \agl{1}{2}\agl{1}{4}\sqr{4}{3}\sqr{5}{4}^2 \\
    && 384\sqrt{2}\pi^2s_{12}^{-3} \agl{1}{2}\agl{1}{5}\sqr{5}{3}\sqr{5}{4}^2 \\
    && 384\sqrt{5}\pi^2s_{12}^{-3} \agl{1}{2}\agl{1}{3}\sqr{4}{3}\sqr{5}{3}\sqr{5}{4} \\
    \midrule
    (-1,0;-1,1,1) & 1 & 192\sqrt{15}\pi^2 s_{12}^{-5/2}\agl{1}{3}^2 \sqr{5}{4}^2\\
    (-1,0;-\hf,\hf,1) & 1 & 192\sqrt{15}\pi^2 s_{12}^{-5/2}\agl{1}{3}^2 \sqr{5}{3} \sqr{5}{4} \\
    && 192\sqrt{15}\pi^2 s_{12}^{-5/2}\agl{1}{3}\agl{1}{4} \sqr{5}{4}^2 \\
    (-1,0;0,0,1) & 1 & 288\sqrt{2}\pi^2s_{12}^{-5/2}\agl{1}{4}^2\sqr{5}{4}^2  \\
    && 288\sqrt{2}\pi^2s_{12}^{-5/2}\agl{1}{3}^2\sqr{5}{3}^2 \\
    && 1152\sqrt{\tfrac{5}{7}}\pi^2s_{12}^{-5/2}\agl{1}{3}\agl{1}{4}\sqr{5}{3}\sqr{5}{4} \\
    (-1,0;0,\hf,\hf) & 1 & 576\pi^2 s_{12}^{-5/2}\agl{1}{4}\agl{1}{5}\sqr{5}{4}^2 \\
    && 576\pi^2 s_{12}^{-5/2}\agl{1}{3}^2\sqr{4}{3}\sqr{5}{3} \\
    && 288\sqrt{10}\pi^2 s_{12}^{-5/2}\agl{1}{3}\agl{1}{5}\sqr{5}{3}\sqr{5}{4} \\
    && 288\sqrt{10}\pi^2 s_{12}^{-5/2}\agl{1}{3}\agl{1}{4}\sqr{4}{3}\sqr{5}{4} \\
    (-1,0;0,1,1) & 1 & 192\sqrt{14}\pi^2s_{12}^{-3}\agl{1}{4}\agl{1}{5}\sqr{5}{4}^3 \\
    && 576\sqrt{7}\pi^2s_{12}^{-3}\agl{1}{3}\agl{1}{5}\sqr{5}{3}\sqr{5}{4}^2 \\
    && 576\sqrt{7}\pi^2s_{12}^{-3}\agl{1}{3}\agl{1}{4}\sqr{4}{3}\sqr{5}{4}^2 \\
    && 576\sqrt{7}\pi^2s_{12}^{-3}\agl{1}{3}^2\sqr{4}{3}\sqr{5}{3}\sqr{5}{4} \\
    (-1,0;\hf,\hf,1) & 1 & 192\sqrt{21}\pi^2s_{12}^{-3}\agl{1}{4}^2\sqr{4}{3}\sqr{5}{4}^2 \\
    && 192\sqrt{21}\pi^2s_{12}^{-3}\agl{1}{3}^2\sqr{4}{3}\sqr{5}{3}^2 \\
    && 1152\sqrt{\tfrac{7}{5}}\pi^2s_{12}^{-3}\agl{1}{4}\agl{1}{5}\sqr{5}{3}\sqr{5}{4}^2\\
    && 1152\sqrt{\tfrac{7}{5}}\pi^2s_{12}^{-3}\agl{1}{3}\agl{1}{5}\sqr{5}{3}^2\sqr{5}{4} \\
    && 1152\sqrt{\tfrac{35}{11}}\pi^2s_{12}^{-3}\agl{1}{3}\agl{1}{4}\sqr{4}{3}\sqr{5}{3}\sqr{5}{4} \\
    (-1,0;1,1,1) & 1 & 384\sqrt{15}\pi^2s_{12}^{-7/2} \agl{1}{5}^2\sqr{5}{3}^2 \sqr{5}{4}^2 \\
    && 384\sqrt{15}\pi^2s_{12}^{-7/2} \agl{1}{4}^2\sqr{4}{3}^2 \sqr{5}{4}^2 \\
    && 384\sqrt{15}\pi^2s_{12}^{-7/2} \agl{1}{3}^2\sqr{4}{3}^2 \sqr{5}{3}^2 \\
    && 576\sqrt{35}\pi^2s_{12}^{-7/2} \agl{1}{4}\agl{1}{5}\sqr{4}{3} \sqr{5}{3}\sqr{5}{4}^2 \\
    && 576\sqrt{35}\pi^2s_{12}^{-7/2} \agl{1}{3}\agl{1}{5}\sqr{4}{3} \sqr{5}{3}^2\sqr{5}{4} \\
    && 576\sqrt{35}\pi^2s_{12}^{-7/2} \agl{1}{3}\agl{1}{4}\sqr{4}{3}^2 \sqr{5}{3}\sqr{5}{4} \\
    \midrule
    (-1,\hf;-1,\hf,1) & \tfrac{3}{2} & 768\pi^2s_{12}^{-5/2} \agl{1}{3}^2\sqr{5}{2}\sqr{5}{4} \\
    (-1,\hf;-\hf,0,1) & \tfrac{3}{2} & 768\sqrt{\tfrac{15}{37}}\pi^2s_{12}^{-5/2} \agl{1}{3}^2\sqr{5}{2}\sqr{5}{3} \\
    && 768\sqrt{\tfrac{15}{13}}\pi^2s_{12}^{-5/2} \agl{1}{3}\agl{1}{4}\sqr{5}{2}\sqr{5}{4} \\
    (-1,\hf;-\hf,\hf,\hf) & \tfrac{3}{2} & 384\sqrt{3}\pi^2s_{12}^{-5/2} \agl{1}{3}^2\sqr{4}{2}\sqr{5}{3} \\
    && 192\sqrt{30}\pi^2s_{12}^{-5/2} \agl{1}{3}\agl{1}{5}\sqr{5}{2}\sqr{5}{4} \\
    && 192\sqrt{30}\pi^2s_{12}^{-5/2} \agl{1}{3}\agl{1}{4}\sqr{4}{2}\sqr{5}{4} \\
    (-1,\hf;-\hf,1,1) & \tfrac{3}{2} & 768\sqrt{\tfrac{21}{5}}\pi^2s_{12}^{-3} \agl{1}{3}^2\sqr{4}{2}\sqr{5}{3}\sqr{5}{4} \\
    && 768\sqrt{\tfrac{21}{5}}\pi^2s_{12}^{-3} \agl{1}{3}\agl{1}{5}\sqr{5}{2}\sqr{5}{4}^2 \\
    && 768\sqrt{\tfrac{21}{5}}\pi^2s_{12}^{-3} \agl{1}{3}\agl{1}{4}\sqr{4}{2}\sqr{5}{4}^2 \\
    (-1,\hf;0,0,\hf) & \tfrac{3}{2} & 768\pi^2s_{12}^{-5/2} \agl{1}{4}^2\sqr{4}{2}\sqr{5}{4} \\
    && 768\sqrt{\tfrac{15}{13}}\pi^2s_{12}^{-5/2} \agl{1}{3}\agl{1}{5}\sqr{5}{2}\sqr{5}{3} \\
    && 768\sqrt{\tfrac{15}{13}}\pi^2s_{12}^{-5/2} \agl{1}{4}\agl{1}{5}\sqr{5}{2}\sqr{5}{4} \\
    && 384\sqrt{10}\pi^2s_{12}^{-5/2} \agl{1}{3}\agl{1}{4}\sqr{4}{2}\sqr{5}{3} \\
    (-1,\hf;0,\hf,1) & \tfrac{3}{2} & 384\sqrt{7}\pi^2s_{12}^{-3} \agl{1}{3}^2\sqr{4}{2}\sqr{5}{3}^2 \\
    && 384\sqrt{7}\pi^2s_{12}^{-3} \agl{1}{4}^2\sqr{4}{2}\sqr{5}{4}^2 \\
    && 768\sqrt{\tfrac{21}{11}}\pi^2s_{12}^{-3} \agl{1}{4}\agl{1}{5}\sqr{5}{2}\sqr{5}{4}^2 \\
    && 384\sqrt{21}\pi^2s_{12}^{-3} \agl{1}{3}\agl{1}{5}\sqr{5}{2}\sqr{5}{3}\sqr{5}{4} \\
    && 768\sqrt{\tfrac{105}{11}}\pi^2s_{12}^{-3} \agl{1}{3}\agl{1}{4}\sqr{4}{2}\sqr{5}{3}\sqr{5}{4} \\
    (-1,\hf;\hf,\hf,\hf) & \tfrac{3}{2} & 384\sqrt{14}\pi^2s_{12}^{-3} \agl{1}{4}^2\sqr{4}{2}\sqr{4}{3}\sqr{5}{4} \\
    && 384\sqrt{14}\pi^2s_{12}^{-3} \agl{1}{5}^2\sqr{5}{2}\sqr{5}{3}\sqr{5}{4} \\
    && 384\sqrt{14}\pi^2s_{12}^{-3} \agl{1}{3}\agl{1}{5}\sqr{4}{2}\sqr{5}{3}^2 \\
    && 384\sqrt{14}\pi^2s_{12}^{-3} \agl{1}{4}\agl{1}{5}\sqr{3}{2}\sqr{5}{4}^2 \\
    && 192\sqrt{105}\pi^2s_{12}^{-3} \agl{1}{3}\agl{1}{4}\sqr{4}{2}\sqr{4}{3}\sqr{5}{3} \\
    && 192\sqrt{105}\pi^2s_{12}^{-3} \agl{1}{4}\agl{1}{5}\sqr{4}{2}\sqr{5}{3}\sqr{5}{4} \\
    (-1,\hf;\hf,1,1) & \tfrac{3}{2} & 512\sqrt{\tfrac{210}{17}}\pi^2s_{12}^{-7/2} \agl{1}{3}^2\sqr{4}{2}\sqr{4}{3}\sqr{5}{3}^2 \\
    && 512\sqrt{15}\pi^2s_{12}^{-7/2} \agl{1}{4}\agl{1}{5}\sqr{3}{2}\sqr{5}{4}^3 \\
    && 1536\sqrt{2}\pi^2s_{12}^{-7/2} \agl{1}{5}^2\sqr{5}{2}\sqr{5}{3}\sqr{5}{4}^2 \\
    && 1536\sqrt{2}\pi^2s_{12}^{-7/2} \agl{1}{4}^2\sqr{4}{2}\sqr{4}{3}\sqr{5}{4}^2 \\
    && 1536\sqrt{\tfrac{42}{13}}\pi^2s_{12}^{-7/2} \agl{1}{4}\agl{1}{5}\sqr{4}{2}\sqr{5}{3}\sqr{5}{4}^2 \\
    && 1536\sqrt{\tfrac{70}{19}}\pi^2s_{12}^{-7/2} \agl{1}{3}\agl{1}{5}\sqr{4}{2}\sqr{5}{3}^2\sqr{5}{4} \\
    && 1536\sqrt{\tfrac{42}{5}}\pi^2s_{12}^{-7/2} \agl{1}{3}\agl{1}{4}\sqr{4}{2}\sqr{4}{3}\sqr{5}{3}\sqr{5}{4}\\
    \midrule
    (-1,1;-1,0,1) & 2 & 96\sqrt{30}\pi^2s_{12}^{-5/2} \agl{1}{3}^2\sqr{5}{2}^2 \\
    (-1,1;-1,\hf,\hf) & 2 & 192\sqrt{15}\pi^2s_{12}^{-5/2} \agl{1}{3}^2\sqr{4}{2}\sqr{5}{2} \\
    (-1,1;-1,1,1) & 2 & 192\sqrt{70}\pi^2s_{12}^{-3} \agl{1}{3}^2\sqr{4}{2}\sqr{5}{2}\sqr{5}{4} \\
    (-1,1;-\hf,0,\hf) & 2 & 960\pi^2s_{12}^{-5/2} \agl{1}{3}\agl{1}{5}\sqr{5}{2}^2 \\
    && 480\sqrt{6}\pi^2s_{12}^{-5/2} \agl{1}{3}\agl{1}{4}\sqr{4}{2}\sqr{5}{2} \\
    (-1,1;-\hf,\hf,1) & 2 & 384\sqrt{\tfrac{105}{11}}\pi^2s_{12}^{-3} \agl{1}{3}^2\sqr{4}{2}\sqr{5}{2}\sqr{5}{3} \\
    && 384\sqrt{21}\pi^2s_{12}^{-3} \agl{1}{3}\agl{1}{5}\sqr{5}{2}^2\sqr{5}{4} \\
    && 192\sqrt{105}\pi^2s_{12}^{-3} \agl{1}{3}\agl{1}{4}\sqr{4}{2}\sqr{5}{2}\sqr{5}{4} \\
    (-1,1;0,0,0) & 2 & 960\pi^2s_{12}^{-5/2} \agl{1}{5}^2\sqr{5}{2}^2 \\
    && 960\pi^2s_{12}^{-5/2} \agl{1}{4}^2\sqr{4}{2}^2 \\
    && 1920\sqrt{\tfrac{3}{7}}\pi^2s_{12}^{-5/2} \agl{1}{4}\agl{1}{5}\sqr{4}{2}\sqr{5}{2} \\
    (-1,1;0,0,1) & 2 & 960\sqrt{\tfrac{21}{11}}\pi^2s_{12}^{-3} \agl{1}{3}\agl{1}{5}\sqr{5}{2}^2\sqr{5}{3} \\
    && 960\sqrt{\tfrac{21}{11}}\pi^2s_{12}^{-3} \agl{1}{4}\agl{1}{5}\sqr{5}{2}^2\sqr{5}{4} \\
    && 960\sqrt{\tfrac{21}{11}}\pi^2s_{12}^{-3} \agl{1}{4}^2\sqr{4}{2}\sqr{5}{2}\sqr{5}{4} \\
    && 480\sqrt{21}\pi^2s_{12}^{-3} \agl{1}{3}\agl{1}{4}\sqr{4}{2}\sqr{5}{2}\sqr{5}{3} \\
    (-1,1;0,\hf,\hf) & 2 & 320\sqrt{21}\pi^2s_{12}^{-3} \agl{1}{5}^2\sqr{5}{2}^2\sqr{5}{4} \\
    && 320\sqrt{21}\pi^2s_{12}^{-3} \agl{1}{4}^2\sqr{4}{2}^2\sqr{5}{4} \\
    && 384\sqrt{35}\pi^2s_{12}^{-3} \agl{1}{3}\agl{1}{4}\sqr{4}{2}^2\sqr{5}{3} \\
    && 192\sqrt{105}\pi^2s_{12}^{-3} \agl{1}{3}\agl{1}{5}\sqr{4}{2}\sqr{5}{2}\sqr{5}{3} \\
    && 640\sqrt{7}\pi^2s_{12}^{-3} \agl{1}{4}\agl{1}{5}\sqr{4}{2}\sqr{5}{2}\sqr{5}{4} \\
    (-1,1;0,1,1) & 2 & 1920\pi^2s_{12}^{-7/2} \agl{1}{3}^2\sqr{4}{2}^2\sqr{5}{3}^2 \\
    && 1920\pi^2s_{12}^{-7/2} \agl{1}{5}^2\sqr{5}{2}^2\sqr{5}{4}^2 \\
    && 1920\pi^2s_{12}^{-7/2} \agl{1}{4}^2\sqr{4}{2}^2\sqr{5}{4}^2 \\
    && 960\sqrt{21}\pi^2s_{12}^{-7/2} \agl{1}{3}\agl{1}{4}\sqr{4}{2}^2\sqr{5}{3}\sqr{5}{4}\\
    && 384\sqrt{30}\pi^2s_{12}^{-7/2} \agl{1}{4}\agl{1}{5}\sqr{4}{2}\sqr{5}{2}\sqr{5}{4}^2 \\
    && 1920\sqrt{\tfrac{42}{11}}\pi^2s_{12}^{-7/2} \agl{1}{3}\agl{1}{5}\sqr{4}{2}\sqr{5}{2}\sqr{5}{3}\sqr{5}{4} \\
    (-1,1;\hf,\hf,1) & 2 & 640\sqrt{42}\pi^2s_{12}^{-7/2} \agl{1}{3}\agl{1}{4}\sqr{4}{2}^2\sqr{5}{3}^2 \\
    && 640\sqrt{42}\pi^2s_{12}^{-7/2} \agl{1}{4}^2\sqr{4}{2}^2\sqr{5}{3}\sqr{5}{4} \\
    && 256\sqrt{105}\pi^2s_{12}^{-7/2} \agl{1}{4}^2\sqr{3}{2}\sqr{4}{2}\sqr{5}{4}^2 \\
    && 1920\sqrt{2}\pi^2s_{12}^{-7/2} \agl{1}{5}^2\sqr{5}{2}^2\sqr{5}{3}\sqr{5}{4} \\
    && 3840\sqrt{\tfrac{7}{13}}\pi^2s_{12}^{-7/2} \agl{1}{3}\agl{1}{5}\sqr{4}{2}\sqr{5}{2}\sqr{5}{3}^2 \\
    && 3840\sqrt{\tfrac{7}{13}}\pi^2s_{12}^{-7/2} \agl{1}{4}\agl{1}{5}\sqr{3}{2}\sqr{5}{2}\sqr{5}{4}^2 \\
    && 3840\sqrt{\tfrac{21}{19}}\pi^2s_{12}^{-7/2} \agl{1}{4}\agl{1}{5}\sqr{4}{2}\sqr{5}{2}\sqr{5}{3}\sqr{5}{4} \\
    (-1,1;1,1,1) & 2 & 288\sqrt{2}\pi^2s_{12}^{-4} \agl{1}{3}\agl{1}{5}\sqr{4}{2}^2\sqr{5}{3}^3 \\
    && 288\sqrt{2}\pi^2s_{12}^{-4} \agl{1}{4}\agl{1}{5}\sqr{3}{2}^2\sqr{5}{4}^3 \\
    && 1152\sqrt{35}\pi^2s_{12}^{-4} \agl{1}{3}\agl{1}{4}\sqr{4}{2}^2\sqr{4}{3}\sqr{5}{3}^2 \\
    && 1152\sqrt{35}\pi^2s_{12}^{-4} \agl{1}{4}\agl{1}{5}\sqr{4}{2}^2\sqr{5}{3}^2\sqr{5}{4} \\
    && 5760\sqrt{2}\pi^2s_{12}^{-4} \agl{1}{4}^2\sqr{4}{2}^2\sqr{4}{3}\sqr{5}{3}\sqr{5}{4} \\
    && 1920\sqrt{6}\pi^2s_{12}^{-4} \agl{1}{4}^2\sqr{3}{2}\sqr{4}{2}\sqr{4}{3}\sqr{5}{4}^2 \\
    && 1920\sqrt{6}\pi^2s_{12}^{-4} \agl{1}{5}^2\sqr{3}{2}\sqr{5}{2}\sqr{5}{3}\sqr{5}{4}^2 \\
    && 1920\sqrt{6}\pi^2s_{12}^{-4} \agl{1}{5}^2\sqr{4}{2}\sqr{5}{2}\sqr{5}{3}^2\sqr{5}{4} \\
    && 960\sqrt{42}\pi^2s_{12}^{-4} \agl{1}{4}\agl{1}{5}\sqr{3}{2}\sqr{4}{2}\sqr{5}{3}\sqr{5}{4}^2 \\
    \midrule
    (-\hf,-\hf;-1,1,1) & 1 & 192\sqrt{30}\pi^2s_{12}^{-5/2} \agl{1}{3}\agl{2}{3}\sqr{5}{4}^2 \\
    (-\hf,-\hf;-\hf,\hf,1) & 0 & 64\sqrt{15}\pi^2s_{12}^{-5/2} \agl{1}{2}\agl{3}{4}\sqr{5}{4}^2 \\
    (-\hf,-\hf;0,0,1) & 0 & 64\sqrt{30}\pi^2s_{12}^{-5/2} \agl{1}{2}\agl{3}{4}\sqr{5}{3}\sqr{5}{4} \\
    (-\hf,-\hf;0,\hf,\hf) & 0 & 32\sqrt{6}\pi^2s_{12}^{-3/2} \agl{1}{2}\sqr{5}{4} \\
    (-\hf,-\hf;0,1,1) & 0 & 64\sqrt{3}\pi^2s_{12}^{-2} \agl{1}{2}\sqr{5}{4}^2 \\
    (-\hf,-\hf;\hf,\hf,1) & 0 & 64\sqrt{6}\pi^2s_{12}^{-2} \agl{1}{2}\sqr{5}{3}\sqr{5}{4} \\
    (-\hf,-\hf;1,1,1) & 0 & 64\sqrt{30}\pi^2s_{12}^{-5/2} \agl{1}{2}\sqr{4}{3}\sqr{5}{3}\sqr{5}{4} \\
    \midrule
    (-\tfrac{1}{2},0;-1,\tfrac{1}{2},1) & \tfrac{1}{2} & 128\sqrt{30}\pi^2 s_{12}^{-5/2}\agl{1}{3}\agl{3}{4}\sqr{5}{4}^2 \\
    (-\tfrac{1}{2},0;-\tfrac{1}{2},0,1) & \tfrac{1}{2} & 384\sqrt{5}\pi^2 s_{12}^{-5/2} \agl{1}{3}\agl{3}{4}\sqr{5}{3}\sqr{5}{4} \\
    && 128\sqrt{10}\pi^2 s_{12}^{-5/2} \agl{1}{2}\agl{3}{4}\sqr{5}{2}\sqr{5}{4} \\
    (-\tfrac{1}{2},0;-\tfrac{1}{2},\tfrac{1}{2},\tfrac{1}{2}) & \tfrac{1}{2} & 128\sqrt{3}\pi^2 s_{12}^{-3/2}\agl{1}{3}\sqr{5}{4} \\
    (-\tfrac{1}{2},0;-\tfrac{1}{2},1,1) & \tfrac{1}{2} & 64\sqrt{30}\pi^2 s_{12}^{-2}\agl{1}{3}\sqr{5}{4}^2 \\
    (-\tfrac{1}{2},0;0,0,\tfrac{1}{2}) & \tfrac{1}{2} & 128\sqrt{2}\pi^2s_{12}^{-3/2}\agl{1}{3}\sqr{5}{3} \\
    && 128\sqrt{2}\pi^2s_{12}^{-3/2}\agl{1}{4}\sqr{5}{4}\\
    (-\tfrac{1}{2},0;0,\tfrac{1}{2},1) & \tfrac{1}{2} & 64\sqrt{15}\pi^2s_{12}^{-2}\agl{1}{4}\sqr{5}{4}^2 \\
    && 128\sqrt{10}\pi^2 s_{12}^{-2}\agl{1}{3}\sqr{5}{3} \sqr{5}{4} \\
    (-\tfrac{1}{2},0;\tfrac{1}{2},\tfrac{1}{2},\tfrac{1}{2}) & \tfrac{1}{2} & 64\sqrt{30}\pi^2 s_{12}^{-2} \agl{1}{3}\sqr{4}{3}\sqr{5}{3} \\
    && 64\sqrt{30}\pi^2 s_{12}^{-2} \agl{1}{4}\sqr{4}{3}\sqr{5}{4} \\
    && 64\sqrt{30}\pi^2 s_{12}^{-2} \agl{1}{5}\sqr{5}{3}\sqr{5}{4} \\
    (-\tfrac{1}{2},0;\tfrac{1}{2},1,1) & \tfrac{1}{2} & 384\sqrt{5}\pi^2 s_{12}^{-5/2} \agl{1}{3}\sqr{4}{3}\sqr{5}{3}\sqr{5}{4} \\
    && 384\sqrt{5}\pi^2 s_{12}^{-5/2} \agl{1}{4}\sqr{4}{3}\sqr{5}{4}^2 \\
    && 384\sqrt{5}\pi^2 s_{12}^{-5/2} \agl{1}{5}\sqr{5}{3}\sqr{5}{4}^2 \\
    \midrule
    (-\hf,\hf;-1,0,1) & 1 & 288\sqrt{10}\pi^2s_{12}^{-5/2} \agl{1}{3}\agl{3}{4}\sqr{5}{2}\sqr{5}{4} \\
    (-\hf,\hf;-1,\hf,\hf) & 1 & 1152\sqrt{\tfrac{5}{13}}\pi^2s_{12}^{-5/2} \agl{1}{3}\agl{3}{5}\sqr{5}{2}\sqr{5}{4} \\
    && 1152\sqrt{\tfrac{5}{13}}\pi^2s_{12}^{-5/2} \agl{1}{3}\agl{3}{4}\sqr{4}{2}\sqr{5}{4} \\
    (-\hf,\hf;-1,1,1) & 1 & 1152\sqrt{\tfrac{35}{31}}\pi^2s_{12}^{-3} \agl{1}{3}\agl{3}{5}\sqr{5}{2}\sqr{5}{4}^2 \\
    && 1152\sqrt{\tfrac{35}{31}}\pi^2s_{12}^{-3} \agl{1}{3}\agl{3}{4}\sqr{4}{2}\sqr{5}{4}^2 \\
    (-\hf,\hf;-\hf,0,\hf) & 1 & 128\sqrt{3}\pi^2s_{12}^{-3/2} \agl{1}{3}\sqr{5}{2} \\
    (-\hf,\hf;-\hf,\hf,1) & 1 & 192\sqrt{5}\pi^2s_{12}^{-2} \agl{1}{3}\sqr{5}{2}\sqr{5}{4} \\
    (-\hf,\hf;0,0,0) & 1 & 192\sqrt{2}\pi^2s_{12}^{-3/2} \agl{1}{4}\sqr{4}{2} \\
    && 192\sqrt{2}\pi^2s_{12}^{-3/2} \agl{1}{5}\sqr{5}{2} \\
    (-\hf,\hf;0,0,1) & 1 & 384\sqrt{\tfrac{5}{7}}\pi^2s_{12}^{-2} \agl{1}{3}\sqr{5}{2}\sqr{5}{3} \\
    && 384\sqrt{\tfrac{5}{7}}\pi^2s_{12}^{-2} \agl{1}{4}\sqr{5}{2}\sqr{5}{4} \\
    (-\hf,\hf;0,\hf,\hf) & 1 & 192\sqrt{5}\pi^2s_{12}^{-2} \agl{1}{3}\sqr{4}{2}\sqr{5}{3} \\
    && 192\sqrt{5}\pi^2s_{12}^{-2} \agl{1}{4}\sqr{4}{2}\sqr{5}{4} \\
    && 192\sqrt{5}\pi^2s_{12}^{-2} \agl{1}{5}\sqr{5}{2}\sqr{5}{4}  \\
    (-\hf,\hf;0,1,1) & 1 & 576\pi^2s_{12}^{-5/2} \agl{1}{5}\sqr{5}{2}\sqr{5}{4}^2  \\
    && 576\pi^2s_{12}^{-5/2} \agl{1}{4}\sqr{4}{2}\sqr{5}{4}^2 \\
    && 288\sqrt{10}\pi^2s_{12}^{-5/2} \agl{1}{3}\sqr{4}{2}\sqr{5}{3}\sqr{5}{4} \\
    (-\hf,\hf;\hf,\hf,1) & 1 & 384\sqrt{3}\pi^2s_{12}^{-5/2} \agl{1}{4}\sqr{3}{2}\sqr{5}{4}^2 \\
    && 384\sqrt{3}\pi^2s_{12}^{-5/2} \agl{1}{3}\sqr{4}{2}\sqr{5}{3}^2 \\
    && 576\sqrt{2}\pi^2s_{12}^{-5/2} \agl{1}{5}\sqr{5}{2}\sqr{5}{3}\sqr{5}{4} \\
    && 192\sqrt{30}\pi^2s_{12}^{-5/2} \agl{1}{4}\sqr{4}{2}\sqr{5}{3}\sqr{5}{4} \\
    (-\hf,\hf;1,1,1) & 1 & 1152\sqrt{\tfrac{35}{31}}\pi^2s_{12}^{-3} \agl{1}{4}\sqr{3}{2}\sqr{4}{3}\sqr{5}{4}^2 \\
    && 1152\sqrt{\tfrac{35}{31}}\pi^2s_{12}^{-3} \agl{1}{5}\sqr{3}{2}\sqr{5}{3}\sqr{5}{4}^2 \\
    && 1152\sqrt{\tfrac{35}{31}}\pi^2s_{12}^{-3} \agl{1}{3}\sqr{4}{2}\sqr{4}{3}\sqr{5}{3}^2 \\
    && 1152\sqrt{\tfrac{35}{31}}\pi^2s_{12}^{-3} \agl{1}{5}\sqr{4}{2}\sqr{5}{3}^2\sqr{5}{4} \\
    && 576\sqrt{14}\pi^2s_{12}^{-3} \agl{1}{4}\sqr{4}{2}\sqr{4}{3}\sqr{5}{3}\sqr{5}{4} \\
    \midrule
    (-\hf,1;-1,\hf,1) & \tfrac{3}{2} & 768\sqrt{\tfrac{21}{11}}\pi^2s_{12}^{-3} \agl{1}{3}\agl{3}{5}\sqr{5}{2}^2\sqr{5}{4}\\
    && 384\sqrt{21}\pi^2s_{12}^{-3} \agl{1}{3}\agl{3}{4}\sqr{4}{2}\sqr{5}{2}\sqr{5}{4} \\
    (-\hf,1;-\hf,0,1) & \tfrac{3}{2} & 64\sqrt{30}\pi^2s_{12}^{-2} \agl{1}{3}\sqr{5}{2}^2 \\
    (-\hf,1;-\hf,\hf,\hf) & \tfrac{3}{2} & 128\sqrt{15}\pi^2s_{12}^{-2} \agl{1}{3}\sqr{4}{2}\sqr{5}{2} \\
    (-\hf,1;-\hf,1,1) & \tfrac{3}{2} & 384\sqrt{6}\pi^2s_{12}^{-5/2} \agl{1}{3}\sqr{4}{2}\sqr{5}{2}\sqr{5}{4} \\
    (-\hf,1;0,0,\hf) & \tfrac{3}{2} & 128\sqrt{15}\pi^2s_{12}^{-2} \agl{1}{5}\sqr{5}{2}^2 \\
    && 256\sqrt{5}\pi^2s_{12}^{-2} \agl{1}{4}\sqr{4}{2}\sqr{5}{2} \\
    (-\hf,1;0,\hf,1) & \tfrac{3}{2} & 768\pi^2s_{12}^{-5/2} \agl{1}{5}\sqr{5}{2}^2\sqr{5}{4} \\
    && 768\sqrt{\tfrac{15}{13}}\pi^2s_{12}^{-5/2} \agl{1}{3}\sqr{4}{2}\sqr{5}{2}\sqr{5}{3} \\
    && 768\sqrt{\tfrac{15}{13}}\pi^2s_{12}^{-5/2} \agl{1}{4}\sqr{4}{2}\sqr{5}{2}\sqr{5}{4} \\
    (-\hf,1;\hf,\hf,\hf) & \tfrac{3}{2} & 192\sqrt{30}\pi^2s_{12}^{-5/2} \agl{1}{5}\sqr{4}{2}\sqr{5}{2}\sqr{5}{3} \\
    && 192\sqrt{30}\pi^2s_{12}^{-5/2} \agl{1}{4}\sqr{3}{2}\sqr{4}{2}\sqr{5}{4} \\
    && 192\sqrt{30}\pi^2s_{12}^{-5/2} \agl{1}{5}\sqr{3}{2}\sqr{5}{2}\sqr{5}{4} \\
    && 384\sqrt{10}\pi^2s_{12}^{-5/2} \agl{1}{4}\sqr{4}{2}^2\sqr{5}{3} \\
    (-\hf,1;\hf,1,1) & \tfrac{3}{2} & 128\sqrt{105}\pi^2s_{12}^{-3} \agl{1}{3}\sqr{4}{2}^2\sqr{5}{3}^2 \\
    && 768\sqrt{\tfrac{21}{5}}\pi^2s_{12}^{-3} \agl{1}{4}\sqr{3}{2}\sqr{4}{2}\sqr{5}{4}^2 \\
    && 768\sqrt{\tfrac{21}{5}}\pi^2s_{12}^{-3} \agl{1}{5}\sqr{3}{2}\sqr{5}{2}\sqr{5}{4}^2 \\
    && 768\sqrt{7}\pi^2s_{12}^{-3} \agl{1}{4}\sqr{4}{2}^2\sqr{5}{3}\sqr{5}{4} \\
    && 384\sqrt{21}\pi^2s_{12}^{-3} \agl{1}{5}\sqr{4}{2}\sqr{5}{2}\sqr{5}{3}\sqr{5}{4} \\
    \midrule
    (0,0;0,0,0) & 0 & 32\sqrt{2}\pi^2 s_{12}^{-1/2}\\
    (0,0;\tfrac{1}{2},\tfrac{1}{2},0) & 0
    & 32\sqrt{6}\pi^2 s_{12}^{-1}\sqr{4}{3}\\
    (0,0;\tfrac{1}{2},\tfrac{1}{2},1) & 0 & 64\sqrt{6}\pi^2 s_{12}^{-3/2}\sqr{5}{3}\sqr{5}{4}\\
    (0,0;\tfrac{1}{2},\tfrac{1}{2},-1) & 0 & 192\sqrt{10}\pi^2s_{12}^{-5/2}\agl{3}{5}\agl{4}{5}\sqr{4}{3}^2 \\
    (0,0;\tfrac{1}{2},-\tfrac{1}{2},0) & 0 & 64 \sqrt{6}\pi^2 s_{12}^{-3/2}\agl{4}{5}\sqr{5}{3} \\
    (0,0;\tfrac{1}{2},-\tfrac{1}{2},1) & 0 & 64 \sqrt{15}\pi^2s_{12}^{-2}\agl{3}{4}\sqr{5}{3}^2 \\
    (0,0;1,1,0) & 0 & 64\sqrt{3}\pi^2s_{12}^{-3/2}\sqr{4}{3}^2 \\
    (0,0;1,1,1) & 0 & 64\sqrt{30}\pi^2s_{12}^{-2}\sqr{4}{3}\sqr{5}{3}\sqr{5}{4} \\
    (0,0;1,1,-1) & 0 & 64\sqrt{210}\pi^2 s_{12}^{-3}\agl{3}{5}\agl{4}{5}\sqr{4}{3}^3 \\
    (0,0;1,-1,0) & 0 & 192\sqrt{5}\pi^2 s_{12}^{-5/2}\agl{4}{5}^2\sqr{5}{3}^2 \\
    (0,0;1,0,0) & 0 & 64 \sqrt{30}\pi^2s_{12}^{-2}\agl{4}{5}\sqr{4}{3}\sqr{5}{3} \\
    \midrule
    (0,\tfrac{1}{2};-1,\tfrac{1}{2},1) & \tfrac{1}{2} & 64\sqrt{210}\pi^2s_{12}^{-3}\agl{3}{4}^2\sqr{4}{2}\sqr{5}{4}^2 \\
    && 128\sqrt{30}\pi^2 s_{12}^{-3}\agl{1}{3}\agl{3}{4}\sqr{2}{1}\sqr{5}{4}^2 \\
    (0,\tfrac{1}{2};-\tfrac{1}{2},0,1) & \tfrac{1}{2} & 128\sqrt{10}\pi^2s_{12}^{-2}\agl{3}{4}\sqr{5}{2}\sqr{5}{4} \\
    (0,\tfrac{1}{2};-\tfrac{1}{2},\tfrac{1}{2},\tfrac{1}{2}) & \tfrac{1}{2} & 64\sqrt{30}\pi^2 s_{12}^{-2} \agl{3}{4}\sqr{4}{2}\sqr{5}{4} \\
    && 128\sqrt{3}\pi^2 s_{12}^{-2}\agl{1}{3}\sqr{2}{1}\sqr{5}{4}\\
    (0,\tfrac{1}{2};-\tfrac{1}{2},1,1) & \tfrac{1}{2} & 384\sqrt{2}\pi^2s_{12}^{-5/2}\agl{3}{4}\sqr{4}{2}\sqr{5}{4}^2 \\
    && 64 \sqrt{30}\pi^2 s_{12}^{-5/2} \agl{1}{3}\sqr{2}{1}\sqr{5}{4}^2\\
    (0,\tfrac{1}{2};0,0,\tfrac{1}{2}) & \tfrac{1}{2} & 64\sqrt{3}\pi^2 s_{12}^{-1}\sqr{5}{2} \\
    (0,\tfrac{1}{2};0,\tfrac{1}{2},1) & \tfrac{1}{2} & 128\sqrt{2}\pi^2s_{12}^{-3/2}\sqr{5}{2}\sqr{5}{4} \\
    (0,\tfrac{1}{2};\tfrac{1}{2},\tfrac{1}{2},\tfrac{1}{2}) & \tfrac{1}{2} & 128\sqrt{3}\pi^2 s_{12}^{-3/2}\sqr{3}{2}\sqr{5}{4} \\
    && 128\sqrt{3}\pi^2 s_{12}^{-3/2}\sqr{4}{3}\sqr{5}{2} \\
    (0,\tfrac{1}{2};\tfrac{1}{2},1,1) & \tfrac{1}{2} & 64\sqrt{30}\pi^2 s_{12}^{-2}\sqr{3}{2}\sqr{5}{4}^2 \\
    && 128\sqrt{10}\pi^2 s_{12}^{-2} \sqr{4}{3}\sqr{5}{2}\sqr{5}{4}\\
    \midrule
    (0,1;-1,0,1) & 1 & 576\sqrt{7}\pi^2 s_{12}^{-3}\agl{3}{4}^2\sqr{4}{2}\sqr{5}{2}\sqr{5}{4} \\
    && 576\sqrt{7}\pi^2 s_{12}^{-3}\agl{3}{4}\agl{3}{5}\sqr{5}{2}^2\sqr{5}{4} \\
    (0,1;-1,\hf,\hf) & 1 & 1152\sqrt{\tfrac{35}{11}}\pi^2 s_{12}^{-3}\agl{3}{4}\agl{3}{5}\sqr{4}{2}\sqr{5}{2}\sqr{5}{4} \\
    && 192\sqrt{21}\pi^2 s_{12}^{-3}\agl{3}{5}^2\sqr{5}{2}^2\sqr{5}{4} \\
    && 192\sqrt{21}\pi^2 s_{12}^{-3}\agl{3}{4}^2\sqr{4}{2}^2\sqr{5}{4} \\
    (0,1;-1,1,1) & 1 & 576\sqrt{35}\pi^2 s_{12}^{-7/2} \agl{3}{4}\agl{3}{5}\sqr{4}{2}\sqr{5}{2}\sqr{5}{4}^2 \\
    && 384\sqrt{15}\pi^2 s_{12}^{-7/2} \agl{3}{5}^2\sqr{5}{2}^2\sqr{5}{4}^2 \\
    && 384\sqrt{15}\pi^2 s_{12}^{-7/2} \agl{3}{4}^2\sqr{4}{2}^2\sqr{5}{4}^2 \\
    (0,1;-\hf,-\hf,1) & 1 & 192\sqrt{5}\pi^2 s_{12}^{-2}\agl{3}{4}\sqr{5}{2}^2 \\
    (0,1;-\hf,0,\hf) & 1 & 128\sqrt{15}\pi^2 s_{12}^{-2}\agl{3}{4}\sqr{4}{2}\sqr{5}{2} \\
    && 96\sqrt{10}\pi^2 s_{12}^{-2}\agl{3}{5}\sqr{5}{2}^2 \\
    (0,1;-\hf,\hf,1) & 1 & 576\pi^2 s_{12}^{-5/2}\agl{3}{5}\sqr{5}{2}^2\sqr{5}{4} \\
    && 288\sqrt{10}\pi^2 s_{12}^{-5/2}\agl{3}{4}\sqr{4}{2}\sqr{5}{2}\sqr{5}{4} \\
    (0,1;0,0,0) & 1 & 192\sqrt{5}\pi^2 s_{12}^{-2} \agl{4}{5}\sqr{4}{2}\sqr{5}{2} \\
    && 192\sqrt{5}\pi^2 s_{12}^{-2} \agl{3}{5}\sqr{3}{2}\sqr{5}{2} \\
    && 192\sqrt{5}\pi^2 s_{12}^{-2} \agl{3}{4}\sqr{3}{2}\sqr{4}{2} \\
    (0,1;0,0,1) & 1 & 192\pi^2 s_{12}^{-3/2}\sqr{5}{2}^2 \\
    (0,1;0,\hf,\hf) & 1 & 192\sqrt{2}\pi^2 s_{12}^{-3/2}\sqr{4}{2}\sqr{5}{2} \\
    (0,1;0,1,1) & 1 & 192\sqrt{5}\pi^2 s_{12}^{-2}\sqr{4}{2}\sqr{5}{2}\sqr{5}{4} \\
    (0,1;\hf,\hf,1) & 1 & 128\sqrt{15}\pi^2 s_{12}^{-2}\sqr{4}{2}\sqr{5}{2}\sqr{5}{3} \\
    && 128\sqrt{15}\pi^2 s_{12}^{-2}\sqr{3}{2}\sqr{5}{2}\sqr{5}{4} \\
    (0,1;1,1,1) & 1 & 192\sqrt{15}\pi^2 s_{12}^{-5/2} \sqr{3}{2}^2\sqr{5}{4}^2 \\
    &&  192\sqrt{15}\pi^2 s_{12}^{-5/2} \sqr{4}{2}^2\sqr{5}{3}^2 \\
    &&  1152\sqrt{\tfrac{5}{7}}\pi^2 s_{12}^{-5/2} \sqr{3}{2}\sqr{4}{2}\sqr{5}{3}\sqr{5}{4} \\
    \midrule
    (\hf,\hf;-1,0,1) & 0 & 192\sqrt{5}\pi^2 s_{12}^{-3}\agl{3}{4}^2\sqr{2}{1}\sqr{5}{4}^2 \\
    (\hf,\hf;-1,\hf,\hf) & 0 & 192\sqrt{10}\pi^2 s_{12}^{-3}\agl{3}{4}\agl{3}{5}\sqr{2}{1}\sqr{5}{4}^2 \\
    (\hf,\hf;-1,1,1) & 0 & 64\sqrt{210}\pi^2 s_{12}^{-7/2}\agl{3}{4}\agl{3}{5}\sqr{2}{1}\sqr{5}{4}^3 \\
    (\hf,\hf;-\hf,-\hf,1) & 1 & 192\sqrt{10}\pi^2 s_{12}^{-2}\agl{3}{4}\sqr{5}{1}\sqr{5}{2} \\
    (\hf,\hf;-\hf,0,\hf) & 0 & 64\sqrt{6}\pi^2 s_{12}^{-2}\agl{3}{4}\sqr{2}{1}\sqr{5}{4} \\
    (\hf,\hf;-\hf,\hf,1) & 0 & 64\sqrt{15}\pi^2s_{12}^{-5/2}\agl{3}{4}\sqr{2}{1}\sqr{5}{4}^2 \\
    (\hf,\hf;0,0,0) & 0 & 32\sqrt{2}\pi^2 s_{12}^{-1}\sqr{2}{1} \\
    (\hf,\hf;0,0,1) & 1 & 192\sqrt{2}\pi^2s_{12}^{-3/2}\sqr{5}{1}\sqr{5}{2} \\
    (\hf,\hf;0,\hf,\hf) & 0 & 32\sqrt{6}\pi^2 s_{12}^{-3/2}\sqr{2}{1}\sqr{5}{4} \\
    (\hf,\hf;0,1,1) & 0 & 64\sqrt{3}\pi^2 s_{12}^{-2}\sqr{2}{1}\sqr{5}{4}^2 \\
    (\hf,\hf;\hf,\hf,1) & 0 & 64\sqrt{6}\pi^2 s_{12}^{-2}\sqr{2}{1}\sqr{5}{3}\sqr{5}{4} \\
    (\hf,\hf;1,1,1) & 0 & 64\sqrt{30}\pi^2 s_{12}^{-5/2}\sqr{2}{1}\sqr{4}{3}\sqr{5}{3}\sqr{5}{4} \\
    \midrule
    (\hf,1;-1,-\hf,1) & \hf & 128\sqrt{30}\pi^2s_{12}^{-3} \agl{3}{4}^2\sqr{2}{1}\sqr{5}{2}\sqr{5}{4} \\
    (\hf,1;-1,0,\hf) & \hf & 384\sqrt{2}\pi^2s_{12}^{-3} \agl{3}{4}^2\sqr{2}{1}\sqr{4}{2}\sqr{5}{4} \\
    && 384\sqrt{5}\pi^2s_{12}^{-3} \agl{3}{4}\agl{3}{5}\sqr{2}{1}\sqr{5}{2}\sqr{5}{4} \\
    (\hf,1;-1,\hf,1) & \hf & 64\sqrt{210}\pi^2s_{12}^{-7/2} \agl{3}{4}^2\sqr{2}{1}\sqr{4}{2}\sqr{5}{4}^2 \\
    && 384\sqrt{14}\pi^2s_{12}^{-7/2} \agl{3}{4}\agl{3}{5}\sqr{2}{1}\sqr{5}{2}\sqr{5}{4}^2 \\
    (\hf,1;-\hf,-\hf,\hf) & \hf & 128\sqrt{3}\pi^2 s_{12}^{-2} \agl{3}{4}\sqr{2}{1}\sqr{5}{2} \\
    (\hf,1;-\hf,0,0) & \hf & 128\sqrt{2}\pi^2 s_{12}^{-2} \agl{3}{4}\sqr{2}{1}\sqr{4}{2} \\
    && 128\sqrt{2} \pi^2s_{12}^{-2} \agl{3}{5}\sqr{2}{1}\sqr{5}{2} \\
    (\hf,1;-\hf,0,1) & \hf & 128\sqrt{10}\pi^2 s_{12}^{-5/2} \agl{3}{4}\sqr{2}{1}\sqr{5}{2}\sqr{5}{4} \\
    (\hf,1;-\hf,\hf,\hf) & \hf & 64\sqrt{30}\pi^2 s_{12}^{-5/2} \agl{3}{4}\sqr{2}{1}\sqr{4}{2}\sqr{5}{4} \\
    && 64\sqrt{30}\pi^2 s_{12}^{-5/2} \agl{3}{5}\sqr{2}{1}\sqr{5}{2}\sqr{5}{4} \\
    (\hf,1;-\hf,1,1) & \hf & 384\sqrt{2}\pi^2 s_{12}^{-3} \agl{3}{4}\sqr{2}{1}\sqr{4}{2}\sqr{5}{4}^2 \\
    && 384\sqrt{2}\pi^2 s_{12}^{-3} \agl{3}{5}\sqr{2}{1}\sqr{5}{2}\sqr{5}{4}^2 \\
    (\hf,1;0,0,\hf) & \hf & 64\sqrt{3}\pi^2 s_{12}^{-3/2} \sqr{2}{1}\sqr{5}{2} \\
    (\hf,1;0,\hf,1) & \hf & 128\sqrt{2}\pi^2 s_{12}^{-2} \sqr{2}{1}\sqr{5}{2}\sqr{5}{4} \\
    (\hf,1;\hf,\hf,\hf) & \hf & 128\sqrt{3}\pi^2 s_{12}^{-2} \sqr{2}{1}\sqr{3}{2}\sqr{5}{4} \\
    && 128\sqrt{3}\pi^2 s_{12}^{-2} \sqr{2}{1}\sqr{4}{2}\sqr{5}{3} \\
    (\hf,1;\hf,1,1) & \hf & 64\sqrt{30}\pi^2 s_{12}^{-5/2} \sqr{2}{1}\sqr{3}{2}\sqr{5}{4}^2 \\
    && 128\sqrt{10}\pi^2 s_{12}^{-5/2} \sqr{2}{1}\sqr{4}{2}\sqr{5}{3}\sqr{5}{4} \\
    \midrule
    (1,1;-1,-1,1) & 1 & 192\sqrt{30}\pi^2s_{12}^{-3} \agl{3}{4}^2\sqr{2}{1}\sqr{5}{1}\sqr{5}{2} \\
    (1,1;-1,-\hf,\hf) & 0 & 64\sqrt{15}\pi^2s_{12}^{-3} \agl{3}{4}^2\sqr{2}{1}^2\sqr{5}{4} \\
    (1,1;-1,0,0) & 0 & 64\sqrt{30}\pi^2s_{12}^{-3} \agl{3}{4}\agl{3}{5}\sqr{2}{1}^2\sqr{5}{4} \\
    (1,1;-1,0,1) & 0 & 192\sqrt{5}\pi^2s_{12}^{-7/2} \agl{3}{4}^2\sqr{2}{1}^2\sqr{5}{4}^2 \\
    (1,1;-1,\hf,\hf) & 0 & 192\sqrt{10}\pi^2s_{12}^{-7/2} \agl{3}{4}\agl{3}{5}\sqr{2}{1}^2\sqr{5}{4}^2 \\
    (1,1;-1,1,1) & 0 & 64\sqrt{210}\pi^2s_{12}^{-4} \agl{3}{4}\agl{3}{5}\sqr{2}{1}^2\sqr{5}{4}^3 \\
    (1,1;-\hf,-\hf,0) & 0 & 32\sqrt{6}\pi^2s_{12}^{-2} \agl{3}{4}\sqr{2}{1}^2 \\
    (1,1;-\hf,-\hf,1) & 1 & 192\sqrt{10}\pi^2s_{12}^{-5/2} \agl{3}{4}\sqr{2}{1}\sqr{5}{1}\sqr{5}{2} \\
    (1,1;-\hf,0,\hf) & 0 & 64\sqrt{6}\pi^2s_{12}^{-5/2} \agl{3}{4}\sqr{2}{1}^2\sqr{5}{4} \\
    (1,1;-\hf,\hf,1) & 0 & 64\sqrt{15}\pi^2s_{12}^{-3} \agl{3}{4}\sqr{2}{1}^2 \sqr{5}{4}^2 \\
    (1,1;0,0,0) & 0 &32\sqrt{2}\pi^2s_{12}^{-3/2} \sqr{2}{1}^2 \\
    (1,1;0,0,1) & 1 & 192\sqrt{2}\pi^2s_{12}^{-2} \sqr{2}{1} \sqr{5}{1}\sqr{5}{2} \\
    (1,1;0,\hf,\hf) & 0 & 32\sqrt{6}\pi^2s_{12}^{-2} \sqr{2}{1}^2 \sqr{5}{4} \\
    (1,1;0,1,1) & 0 & 64\sqrt{3}\pi^2s_{12}^{-5/2} \sqr{2}{1}^2 \sqr{5}{4}^2 \\
    (1,1;\hf,\hf,1) & 0 & 64\sqrt{6}\pi^2s_{12}^{-5/2} \sqr{2}{1}^2 \sqr{5}{3}\sqr{5}{4} \\
    (1,1;1,1,1) & 0 & 64\sqrt{30}\pi^2s_{12}^{-3} \sqr{2}{1}^2 \sqr{4}{3}\sqr{5}{3}\sqr{5}{4} \\
    \bottomrule
    \bottomrule
\end{eqnarray*}}

\end{document}